\documentclass[12pt,preprint]{aastex}
\usepackage{graphics, enumerate, graphicx, color}
\usepackage{psfig, epsfig}
\usepackage{alltt}
\usepackage{natbib}

\newcommand{\bearr}{\begin{eqnarray}}
\newcommand{\eearr}{\end{eqnarray}}
\definecolor{titlecol}{rgb}{0,0,1}
\definecolor{titledark}{rgb}{0,0,0.8}
\definecolor{hilit}{rgb}{0,0,1}
\definecolor{hilitdark}{rgb}{0,0,0.8}

\def\degr               {^{\circ}}

\def\csq			{$\chi^2$}
\def\mcsq		{\chi^2}
\def\csqnu              {$\chi^2_{\nu}$}
\def\mcsqnu             {\chi^2_{\nu}}
\def\sigr		{\sigma_{r_e} / r_e}

\newcommand{\eqnstart}   {\begin{equation}}
\newcommand{\eqnend}     {\end{equation}}

\def\mBTlet	{\left(\mbox{B}/\mbox{Tot}\right)}
\def\BT		{bulge-to-total}
\def\mBT	{(L_B/L_{tot})}
\def\zpr	{z^{\prime}}

\def\lcontr  {\log{\left( L_{\mbox{\small{host}}}/L_{\mbox{\small{PS}}} \right)}}
\def\contr  {\left( L_{\mbox{\small{host}}}/L_{\mbox{\small{PS}}} \right)}
\def\CR		{host-to-PS}

\slugcomment{Accepted for publication in Ap. J.}

\shorttitle{Recovery of AGN Host Galaxy Morphologies}
\shortauthors{Simmons \& Urry}

\begin{document}

\title{The Accuracy of Morphological Decomposition of AGN Host Galaxies}


\author{B. D. Simmons and C. M. Urry}
\affil{Astronomy Department, Yale University, New Haven, CT USA}

\email{simmons@astro.yale.edu}

\begin{abstract}
In order to assess the accuracy with which we can determine the
morphologies of AGN host galaxies, we have simulated more than 50,000
ACS images of galaxies with $z < 1.25$, using image and noise
properties appropriate for the GOODS survey. We test the effect of
central point-source brightness on host galaxy parameter recovery with
a set of simulated AGN host galaxies made by adding point sources to
the centers of normal galaxies. We extend this analysis and also
quantify the recovery of intrinsic morphological parameters of AGN
host galaxies with a set of fully simulated inactive and AGN host
galaxies.

We can reliably separate good from poor fit results using a
combination of reasonable error cuts, in the regime where $\contr >
1:4$. 
We give quantitative estimates of parameter errors as a function
of host-to-point-source ratio.  
In general, we separate host and point-source magnitudes reliably
at all redshifts; point sources are well recovered more than 90\% of
the time, although spurious detection of central point sources can be
as high as 25\% for bulge-dominated sources.
We find a general correlation between S\'ersic index and intrinsic
bulge-to-total ratio, such that a host galaxy with S\'ersic $n < 1.5$
generally has at least 80\% of its light from a disk
component. Likewise, ``bulge-dominated'' galaxies with $n > 4$
typically derive at least 70\% of their total host galaxy light from a
bulge, but this number can be as low as 55\%. Single-component
S\'ersic fits to an AGN host galaxy are statistically very reliable to
$z < 1.25$ (for ACS survey data like ours). In contrast,
two-component fits involving separate bulge and disk components tend
to over-estimate the bulge fraction by $\sim 10$\%, with uncertainty
of order 50\%.

\end{abstract}

\keywords{methods: data analysis --- galaxies: active --- galaxies: nuclei --- galaxies: bulges --- galaxies: fundamental parameters}

\pagebreak

\section{Introduction}
Morphological analysis of two-dimensional light profiles of galaxies
in large data surveys provides detailed information about galaxy
populations and their evolution \citep[\emph{e.g.,}][]{simard99,
ravindranath04, jogee04, sheth08}. Simulations of both large and small
galaxy samples show that two-dimensional parametric and non-parametric
morphological analysis of normal galaxies is extremely robust
\citep{marleau98, graham01a, conselice03, trujillo04, haussler07}.

Studies of active galactic nucleus (AGN) host galaxies give us the
opportunity to study not just the galaxies themselves, but also the
well-established connection between galaxies and central supermassive
black holes \citep{kormendy95, magorrian98, ferrarese00, gebhardt00,
mclure02, marconi03}. In cases where central black hole masses are
independently determined, accurate decompositions of AGN host galaxies
from their central point-source light contributions allow for the
direct study of how galaxy light distribution relates to black hole
properties. Alternatively, when time-intensive observations of black
hole mass are not available, we can use established bulge-black hole
relations to determine the black hole masses from bulge luminosities.

However, two-dimensional morphological fitting of AGN host galaxies is
more complicated than that of ``normal'' (or ``inactive'') galaxies
because of the presence of a central point source, which is often
quite bright. In order to extract host galaxy properties from a source
comprised of a host plus a central AGN, spatial resolution is
critical. Thus the Advanced Camera for Surveys (ACS) on the
\emph{Hubble Space Telescope} (\emph{HST}) is the instrument of choice
for many AGN host galaxy studies to date \citep{sanchez04,
alonsoherrero08, ballo07}.

Of the large multi-wavelength surveys currently available, the Great
Observatories Origins Deep Survey \citep[GOODS;][]{giavalisco04}
provides some of the deepest multicolor ACS data. AGN identification
is possible because of deep X-ray imaging with \emph{Chandra}, as well as
ground-based optical and infrared spectroscopic follow up. \emph{Spitzer}
data provide additional information on total light but insufficient
spatial resolution to separate the galaxy from the active nucleus.  We
performed detailed morphological analysis on the GOODS ACS data
set using GALFIT \citep{peng02}, the results of which are presented in
a forthcoming paper (B. D. Simmons et al. 2008, in preparation, hereafter S08).  In
order to understand the accuracy of those results -- in particular, to
probe the well-known effects of surface brightness dimming and
dependence of physical resolution on redshift in the presence of a
central point source whose size does not change with redshift -- 
extensive simulations of host galaxy morphology are required.

Previously, \citet{sanchez04} simulated 1880 single-component host
galaxies with point sources fainter than the host galaxy.  Here we
present a full treatment of over 50,000 simulated AGN host galaxies in
the redshift range $0.1 < z < 1.1$, with both single-component and
two-component bulge-plus-disk morphologies, and central point sources
that are both brighter and fainter than the host galaxy. Results of
these simulations are intended to inform data analysis of AGN host
morphologies, to better infer intrinsic host galaxy shapes from fitted
morphological parameters in the presence of a central point source.

We discuss the data from which we draw our noise properties and
simulated sample in Section 2. The detailed fit procedures and the
simulation parameter space are presented in Section 3, and we assess
the ability of our fitting procedure to recover accurate host galaxy
and point-source parameters from GOODS-like images of AGN host
galaxies in Section 4.

\section{Data}
\subsection{HST ACS Data}
The GOODS fields each cover an area of approximately $10^{\prime}
\times 16^{\prime}$ with a total of 398 \emph{HST} orbits in both
fields. The ACS has a resolution of $0^{\prime \prime}.05$
pixel$^{-1}$, and observations were taken in the F435W, F606W, F775W,
and F850LP ACS bands, hereafter referred to as $B$, $V$, $I$, and
$\zpr$, respectively. Data acquisition and reduction are detailed in
\citet{giavalisco04} and \citet{koekemoer02}. Each of the five epochs
of data was processed via the basic ACS pipeline and a further
processing task called multidrizzle, which finds an astrometric
solution for all 5 epochs in order to correct for geometric
distortion, and at the same time removes cosmic rays from the image
\citep{koekemoer02}. The final images have resolution of $0^{\prime
\prime}.03$ pixel$^{-1}$, and the magnitude limits for extended sources
are $B < 28.4$ mag, $V < 28.4$ mag, $I < 27.7$ mag, and $z < 27.3$
mag.

From the reduced data, we randomly selected 450 inactive galaxies with
stellarity class less than 0.8 (\emph{i.e.,} objects that are not
point sources) and magnitude $\zpr < 24.0$ mag
for use in those of our simulations that include real data (\S3.3).
Since the majority of AGN host galaxies selected for morphological
analysis in current surveys using \emph{HST} have redshifts 
such that their $I$-band data lie in the rest-frame $B$ band ($0.575 <
z < 0.9$), we performed simulations using the $I$-band images and
noise properties.

\subsection{Noise Properties}
\begin{figure}
\plotone{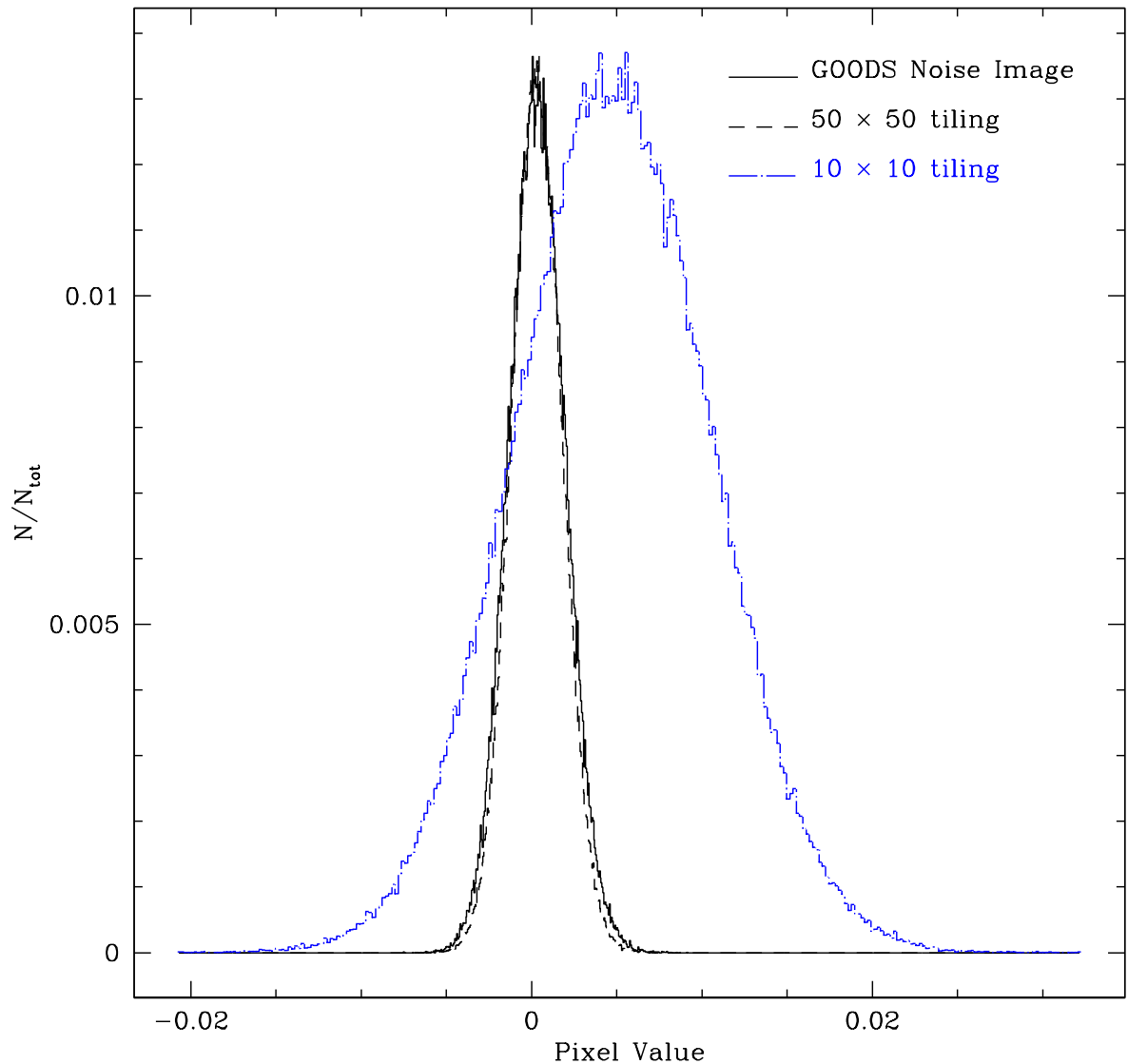}
\caption{
Histograms of pixel intensities for GOODS noise (solid), noise mosaics
using a $50 \times 50$-pixel sampling of the GOODS noise (dashed), and
using a $10 \times 10$-pixel sampling of the GOODS noise (blue, dot-dashed). The
histogram for the mosaic using $10 \times 10$-pixel tiles is
significantly different from the intrinsic noise distribution. The
distribution of pixel intensity of the $50 \times 50$ noise mosaic is
nearly indistinguishable from that of the intrinsic GOODS noise
distribution.   }
\label{noisehist}
\end{figure}

For fully simulated galaxies (described in \S3.4), noise appropriate
to the GOODS-ACS fields was added. The task of simulating the noise in
the GOODS-ACS fields is complicated by the dithering process, which
correlates the noise among nearby pixels. We therefore use actual
noise pixels (i.e., source-free pixels) from the final GOODS-N
and GOODS-S images rather than statistical models. We sampled 150
sub-sections of the GOODS images that contained no sources, 75 in the
North field and 75 in the South, for a total of over 2.5 million noise
pixels. Each of these sub-sections was cut into tiles of size $50
\times 50$ pixels; this resulted in 1042 tiles. These were then used
to create 500 noise images of the same size as our data images by
arranging random mosaics of the $50 \times 50$ noise tiles. We have
verified that these new noise mosaics have the same overall noise
properties (distribution of pixel intensities) as the original
noise-only subsections of the GOODS images (Figure \ref{noisehist}).

We also tested noise mosaics created from $10 \times 10$ tiles of
background noise but found that the distribution of pixel intensities was
shifted compared to the correlated noise properties of the dithered
ACS images. We thus used only the $50 \times 50$ pixel noise images to
create our final noise mosaics. The noise value distributions of the $50
\times 50$ and $10 \times 10$ noise mosaics are compared to the actual 
GOODS noise distribution in Figure \ref{noisehist}.

\section{Creation and Morphological Fitting of Simulated Samples}

We performed two distinct sets of simulations to test the accuracy of
derived morphological parameters. First, we used a sample of 450 real
galaxies from the GOODS fields and added point sources to their
centers. This allowed us to test the ability of GALFIT to extract
point sources from the centers of real galaxies. It also informed our
choice of parameters and galaxy types for the second, more extensive,
set of simulations. In the latter set of simulations, we created a
sample of 12,592 completely simulated GOODS galaxies, in order to
additionally test the recovery of a wide range of host galaxy
parameters. We fitted all of our simulated galaxies with our
batch-fitting algorithm (described below).  We also simulated the
redshifting of these galaxies to quantify redshift-dependent effects,
and we added noise taken directly from real GOODS data.

\subsection{Fit Procedure}
We performed morphological analysis in two dimensions using GALFIT,
which can simultaneously fit an arbitrary number of components to an
image \citep{peng02}. The program uses a \csq\ minimization method to
determine the best-fit parameters. We used the S\' ersic profile,
which models the light distribution of a galaxy as an exponential
function with a variable half-light radius, $r_e$, and an exponential
parameter, $n$, called the S\'ersic index \citep{sersic68}. When $n$
is fixed at a value of 1, the S\'ersic profile is equivalent to an
exponential disk; when $n=4$, the S\'ersic profile is equivalent to a
de Vaucouleurs bulge. Each of our simulated galaxies was fitted twice:
once using a single S\'ersic function with a variable index, and once
with two S\'ersic functions with fixed $n=4$ and $n=1$ (bulge +
disk). We also simultaneously fitted point-source (PS) components
using the point-spread function (PSF) of the data, determination of
which is based on GOODS field stars and is described in S08.

We developed a batch-fitting algorithm to fit AGN host galaxies in
GALFIT (described in detail in S08), which we use here on our
simulated galaxies. The batch-fitting algorithm uses a series of
initial guesses to execute a first-pass fit of the central region of
each AGN host galaxy, fitting the individual centroid positions of
each component (S\'ersic plus point source, or de Vaucouleurs bulge plus
exponential disk plus point source). The second fit iteration zooms
out to include the full extended galaxy and fits the central point-source magnitude and AGN host galaxy parameters (S\'ersic index or
bulge/disk decomposition, half-light radius, axial ratio,
etc.). Finally, a third iteration is performed where all parameters
are allowed to vary (except $n$ in cases of a bulge + disk fit).

It is important to note that, while these simulations include
thousands of AGN host galaxies, even the largest (current) galaxy
surveys using ACS data provide no more than hundreds of AGN host
galaxies that can be reasonably fitted with GALFIT. We can therefore (as
in S08) follow up the batch fits by hand, using the results of the
batch-mode fitting as initial guesses to further constrain the results
for each AGN host galaxy. Due to the size of the simulated galaxy
sample in this paper, we did not do this final hand-fitting step.
Thus our conclusions about the fraction of cases for which accurate
morphological parameters are recovered should be considered
conservative when compared to a sample of galaxies whose fits are
adjusted and verified individually.

\subsection{Determination of Initial Parameter Guesses}
For our real host galaxy sample, we use magnitude, flux radius, and
position angle from the GOODS catalogs, which were created using
SExtractor \citep{bertin96}. Our simulated galaxies do not have a
SExtractor catalog, but we simulated initial parameter guesses by
introducing random errors into the true values.

Specifically, when determining initial guesses for parameters that can
be measured from a data image, we assume that the guess value is accurate
to within the following: $\pm (2.0,2.0)$ pixels in position, $\pm 0.5$
in total magnitude, $10\%$ in $r_e$ and $b/a$, and $10 \degr$ in
position angle. We assumed no \emph{a priori} knowledge of the
S\'ersic index $n$ of each simulated galaxy, nor of whether each
simulated galaxy was a two-component bulge plus disk system, or a
system with either a pure bulge or disk. Our initial guess for the
S\'ersic index (or indices) of each fit is $n=2.5$.

In the case of a multi-component fit, we assume that the total
magnitude is split evenly among the components (i.e., that we have no
prior knowledge of host:PS contrast ratio or \BT\ ratio), and we
additionally assume that the measured position angle and axial ratio
are the average of the actual values for each separate component after
adding the random fluctuation to the sum.

\subsection{Real Galaxies with Added Point Sources}

In order to test whether we can recover central point sources from
galaxies, we selected 450 $I$-band images of normal galaxies (i.e.,
not X-ray detected) from the GOODS-North and GOODS-South fields. The
selection was performed so the sample has the same magnitude
distribution as the full sample of galaxies in each of the GOODS
fields, but was otherwise random. We assumed no knowledge of redshifts
within the sample.

We fitted each of these galaxies with a single S\'ersic profile in order to
determine the baseline set of morphological parameters for each
galaxy. These fit parameters (S\'ersic $n$, magnitude, half-light
radius $r_e$, etc.) were taken to be the ``actual'' galaxy
parameters. 

We then made nine copies of each galaxy, adding one point source to
each, centered on the central pixel of the galaxy. The added point
sources ranged in $I$-band magnitude from 21 to 29, in increments of 1
mag. This gave us a total of 4050 simulated AGN from the original 450
galaxies, with a large range of host galaxy-to-point-source contrast
ratios, from $\sim$1000:1 to 1:100. These provide a straightforward
means of assessing the effect of a central point source on the
recovered parameters of a real galaxy.

\begin{figure}
\plotone{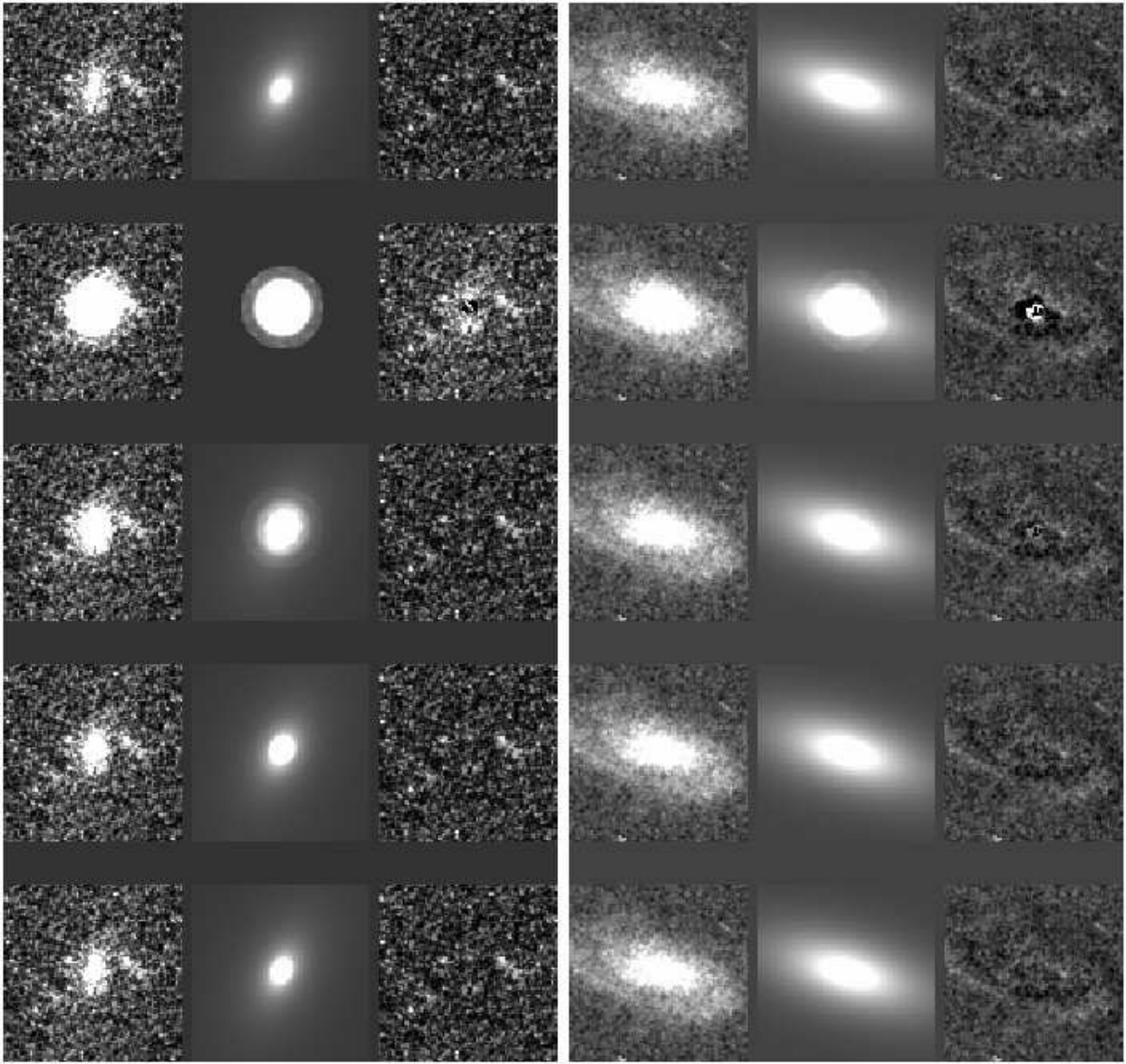}
\caption{Example fit results for two AGN hosts simulated from real
galaxies. For each galaxy, the top row shows the original GOODS galaxy
(from left to right: galaxy, fit, residual). Each row below shows the
host+fit+residual for added central point sources of magnitudes
$I_{AB} = 22$, 24, 26, and 28, with the faintest point source in the
bottom row.   \emph{Note:} These cutouts are zoomed-in to show the galaxies;
the actual fitting regions are significantly larger.}
\label{psfadd_gals}
\end{figure}

Our convergence rate for the fitting routine depends on the magnitude
of the added central point source. Of the models with $I_{PS} = 21$,
88\% converge, whereas 64\% of models with $I_{ps} = 29$
(significantly below the flux limit of our $I$-band sample)
converged. This is due to the fitting program being unable to converge
to a value for the central point-source magnitude. Examples of fit
results for typical galaxies in our sample are shown in Figure
\ref{psfadd_gals}.

\subsection{Simulated Galaxies}

The simulated hosts described in \S3.3 provide information on how the
presence of a central point source changes the fitted host galaxy
parameters, but they do not allow us to explore redshift effects or
quantify the absolute accuracy of the fit parameters. This can only be
accomplished with a set of simulations for which we know the intrinsic
properties of each simulated galaxy.

To that end, we simulated three kinds of galaxies with a range of
parameters: pure de Vaucouleurs bulge (B) galaxies with fixed S\'ersic
$n=4$, pure exponential disk (D) galaxies with $n=1$, and composite
galaxies with both bulge and disk (B+D) components in varying
proportions. The galaxy parameters were chosen to be typical of
bright, local galaxies \citep{binneymerrifield} and placed at
$z=0.125$. Redshifted samples were developed from these, as described
below.

For the B and D type galaxies, the component parameters occupy a grid
of four values for each parameter across the ranges $16.9 \leq I_{AB}
\leq 20.4$ and $1.5 \leq r_e \leq 6.0$ kpc (for bulges); $4.0 \leq r_e
\leq 10.0$ kpc (for disks), with axis ratios $0.25 \leq b/a \leq
1.0$. For the single-component fits, the position angle was fixed at
$45.0 \degr$. This results in 64 B and 64 D galaxies, each of which is
``inactive'', i.e., without a central point source. We create
simulated AGN host galaxies by adding point sources to these. We used
five values between $16.5 \leq I_{AB} \leq 24$ for the central point
source, resulting in 320 B-type and 320 D-type AGN host galaxies, with
contrast ratios ranging from $-1.3 \leq \lcontr \leq 2.8$. Lastly, in
order to ensure enough galaxies for statistical analysis after binning
the fitted single-component sample, we made four copies of each B and
D inactive galaxy and two copies of each B and D AGN host galaxy. This
resulted in a total of 256 B and 256 D inactive single-component
galaxies, and 640 B and 640 D AGN host galaxies.

For the B+D type galaxies, the parameter space includes more
magnitudes so as to provide 20 input \BT\ ratios with $0.028 \leq
\mBTlet \leq 0.97$ [plus the single-component B and D fits, which have
$\mBTlet = 1$ and 0, respectively]. In addition, the position angles
of the individual components were allowed to vary such that some of
the simulated B+D galaxies have components that are slightly ($\leq
15^{\circ}$) off-axis with respect to each other. The axis ratios
between bulges and disks were also allowed to vary with respect to
each other. These changes significantly increase the total number of
double-component galaxies created; thus the number of radius, $b/a$,
and point-source magnitude parameters used was decreased in order to
keep the total number of galaxies from being prohibitively large from
a computational perspective. This resulted in the creation of 2700
inactive double-component (B+D) galaxies, and 8100 B+D AGN host
galaxies. Table \ref{sersic_param_table} gives the parameter values
used to create the entire local suite of 12,592 simulated single- and
double-component galaxies.

Since we are also interested in distinguishing the effects of
redshifting galaxies from the effects of evolving galaxies, the
initial sample was defined to lie at $z= 0.125$, and three additional
samples were placed at $z=0.413, 0.738$, and 1.075, corresponding to
the redshifts at which the centers of each of our GOODS filters
(F435W, F606W, F775W, and F850LP) are in the rest-frame $B$ (F435W)
band.

To redshift each galaxy, we used a concordance cosmology with
$\Omega_{tot} = 1$, $\Omega_{\Lambda} = 0.73$, and $H_0 = 71$ km
s$^{-1}$ Mpc$^{-1}$ \citep{spergel03} to calculate the cosmological
dimming and loss of resolution corresponding to each redshift. Because
the size of the ACS PSF does not change with redshift, we assume that
the size of AGN central point sources will also stay fixed with
redshift, unlike the host galaxy. We therefore created the redshifted
AGN by redshifting the model host galaxy (in flux and size) separately
from the central point source (in flux only) and adding them together
before convolving with the ACS PSF and adding noise to the image.  The
redshifting of each of the 12,592 galaxies produced another 12,592
galaxies at each final redshift, for a total of 50,368 fully simulated
galaxies located at four different redshifts from $0.125 < z < 1.075$.

\begin{figure}
\plotone{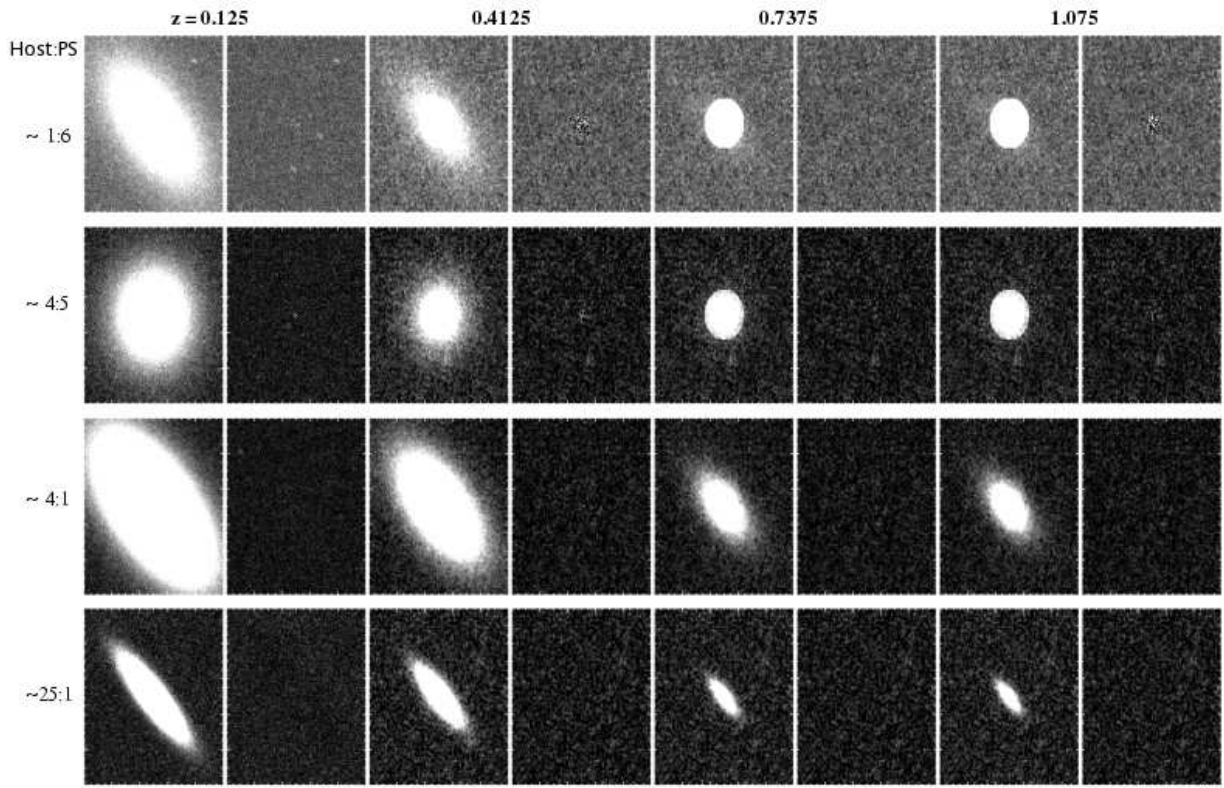}
\caption{Example fit results for four fully simulated galaxies at four
different redshifts. Each row shows one galaxy. The first two columns
show the image of the galaxy at $z=0.125$ and residual. Each
successive pair of columns from left to right shows galaxy and
residual for $z=0.4125, 0.7375$, and 1.075, respectively. \emph{Note:}
These cutouts are zoomed-in to show the galaxies; the actual fitting
regions are significantly larger.}
\label{sim_gals}
\end{figure}

We fit each using our batch-fitting algorithm. Each galaxy was fit
twice: once with a generalized S\'ersic profile plus a point-source
component, and once with a combination of de Vaucouleurs bulge,
exponential disk, and point source. Inactive galaxies were also fit
without point-source components. Examples of fit results for typical
galaxies in our sample are shown in Figure \ref{sim_gals}.

\noindent
\begin{table}[h]
\begin{center}

\begin{tabular}{|l|c|c|c|c|}\hline
\tablecaption{Parameter Grid Values for Simulated Galaxies}
\setlength{\tabcolsep}{0.04in}
\tablecolumns{5}

          & Pure Bulges & \ Pure Disks\    & \multicolumn{2}{c|}{Bulge + Disk} \\
          &             &                  & \em{Bulge}  & \em{Disk} \\ \hline
$m_{gal}$ & 16.898, 5.0 & 16.898, 5.0      & 16.898, 5.0 & 16.898, 5.0    \\
($I_{AB}$, $L_*$)& 18.645, 1.0& 18.645, 1.0& 18.645, 1.0 & 18.645, 1.0    \\
          & 19.398, 0.5 & 19.398, 0.5      & 19.398, 0.5 & 19.398, 0.5    \\
          & 20.393, 0.2 & 20.393, 0.2      & 20.393, 0.2 & 20.393, 0.2    \\
          &             &                  & 20.750, 0.14& 20.750, 0.14   \\ \hline
$r_e$     & 13.54, 1.5     & 36.12, 4.0    & 27.09, 3.0 & 54.18, 6.0   \\
(pixels, kpc) & 27.09, 3.0 & 54.18, 6.0    & 40.63, 4.5 & 72.23, 8.0   \\
          & 40.63, 4.5     & 72.23, 8.0    & 54.18, 6.0 & 90.29, 10.0  \\
          & 54.18, 6.0     & 90.09, 10.0   &            &              \\ \hline
$b/a$     & 0.25        & 0.25             & 0.65             & 0.30      \\
          & 0.50        & 0.50             & 1.0              & 0.65      \\
          & 0.75        & 0.75             &                  & 1.0       \\
          & 1.0         & 1.0              &                  &           \\ \hline
PA        & 45.0        & 45.0             & 45.0             & 45.0      \\
(degrees) &             &                  & 60.0             &           \\ \hline
$m_{PS}$  & none        & none             & \multicolumn{2}{c|}{none  }  \\
($I_{AB}$)& 16.500      & 16.500           & \multicolumn{2}{c|}{16.500}  \\
          & 18.375      & 18.375           & \multicolumn{2}{c|}{20.250}  \\
          & 20.250      & 20.250           & \multicolumn{2}{c|}{22.125}  \\
          & 22.125      & 22.125           & \multicolumn{2}{c|}{ }       \\
          & 24.000      & 24.000           & \multicolumn{2}{c|}{ }       \\ \hline

\tableline
\end{tabular}
\caption{Grid values for simulated galaxies at $z=0.125$. The
effective radius $r_e$ values differ between the bulge and disk
galaxies due to the different physical sizes of these two classes of
galaxies. The parameter space varies slightly between single-component
galaxies and double-component galaxies in order to increase the number
of mesh points in the \BT\ ratio without increasing the simulated
double-component galaxies to a computationally prohibitive number.
Conversion from magnitude to luminosity and between pixels and
physical size uses a concordance cosmology ($\Omega_{tot} = 1$,
$\Omega_{\Lambda} = 0.73$, and $H_0 = 71$ km/s/Mpc) and a redshift $z
= 0.125$. }
\label{sersic_param_table}
\end{center}
\end{table}

\section{Results And Discussion}

\subsection{Existing Galaxies with Added Point Sources}

We are interested mainly in three properties of these models: how well
they recover the input point-source magnitude, how well they recover
the baseline host galaxy parameters, and how the accuracy of the host
galaxy parameters relates to the point-source magnitude and/or the
contrast ratio between central point source and host galaxy.

\subsubsection{AGN (Point Source) Recovery}

\begin{figure}
\plotone{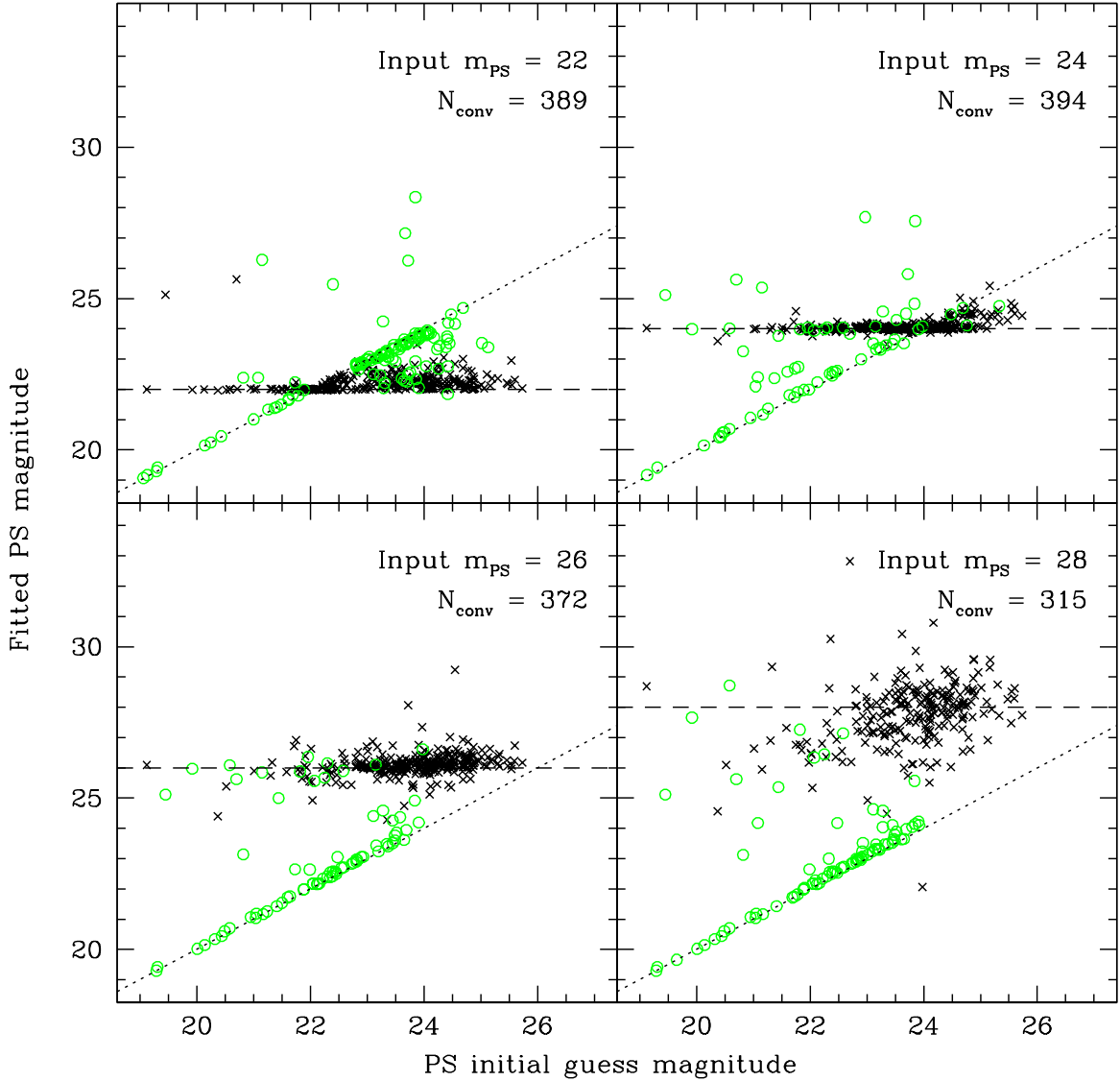}
\caption{ Recovery of magnitude of a point source added to real galaxy
images, for: input point-source magnitudes of 22, 24, 26, and 28 mag.
The fitting recovers the point-source magnitude nicely after making a
simple $\mcsqnu < 2.0$ cut (black crosses), with a larger spread for
fainter values of the input point source, as expected. Fits with a
large value of $\mcsqnu$ are shown as green open circles; $N_{conv}$
indicates the number of fits (out of 450) that converged automatically
for this input point-source magnitude. The dashed horizontal line
shows $m_{\mbox{\tiny out}} = m_{\mbox{\tiny in}}$, so fits falling on
this line perfectly recover the actual point-source magnitude. The
dotted line shows $m_{\mbox{\tiny out}} = m_{\mbox{\tiny guess}}$:
fits on this line converged to their initial guess point-source
magnitude, and the final fitted magnitude is wrong; nearly all of
these fits are removed by the $\mcsqnu$ cut.  }

\label{m2m0_all}
\end{figure}

Figure \ref{m2m0_all} shows the degree of recovery of the central
point-source magnitude for input point sources. As expected, brighter
point sources are more accurately recovered.  At all values of input
point-source magnitude, the fitted value of the point-source magnitude
tends to either converge to the true input magnitude \emph{or} remain
at the input guess magnitude.

In all cases, we can easily separate these two groups with a simple
reduced \csqnu\ cut ($\mcsqnu < 2.0$). After the cut, the number of
remaining poor fits is very low, as is the number of good fits removed
by the cut: both are less than 5\%, even in the limiting cases
of very high and very low input point-source magnitude. This level of
contamination/excess removal of good fits is not strongly dependent on
the value at which we choose to cut $\mcsq$, nor is it strongly
dependent on the input point-source magnitude.

The rms uncertainties of the recovered point-source magnitudes for
each input point-source value are shown in Table
\ref{rms_params}. These values reflect the systematic uncertainties in
the fitted parameters, and should be added to the statistical error in
fitted point-source magnitude returned by GALFIT.

\subsubsection{Host Galaxy Parameter Recovery}

\begin{figure}
\plotone{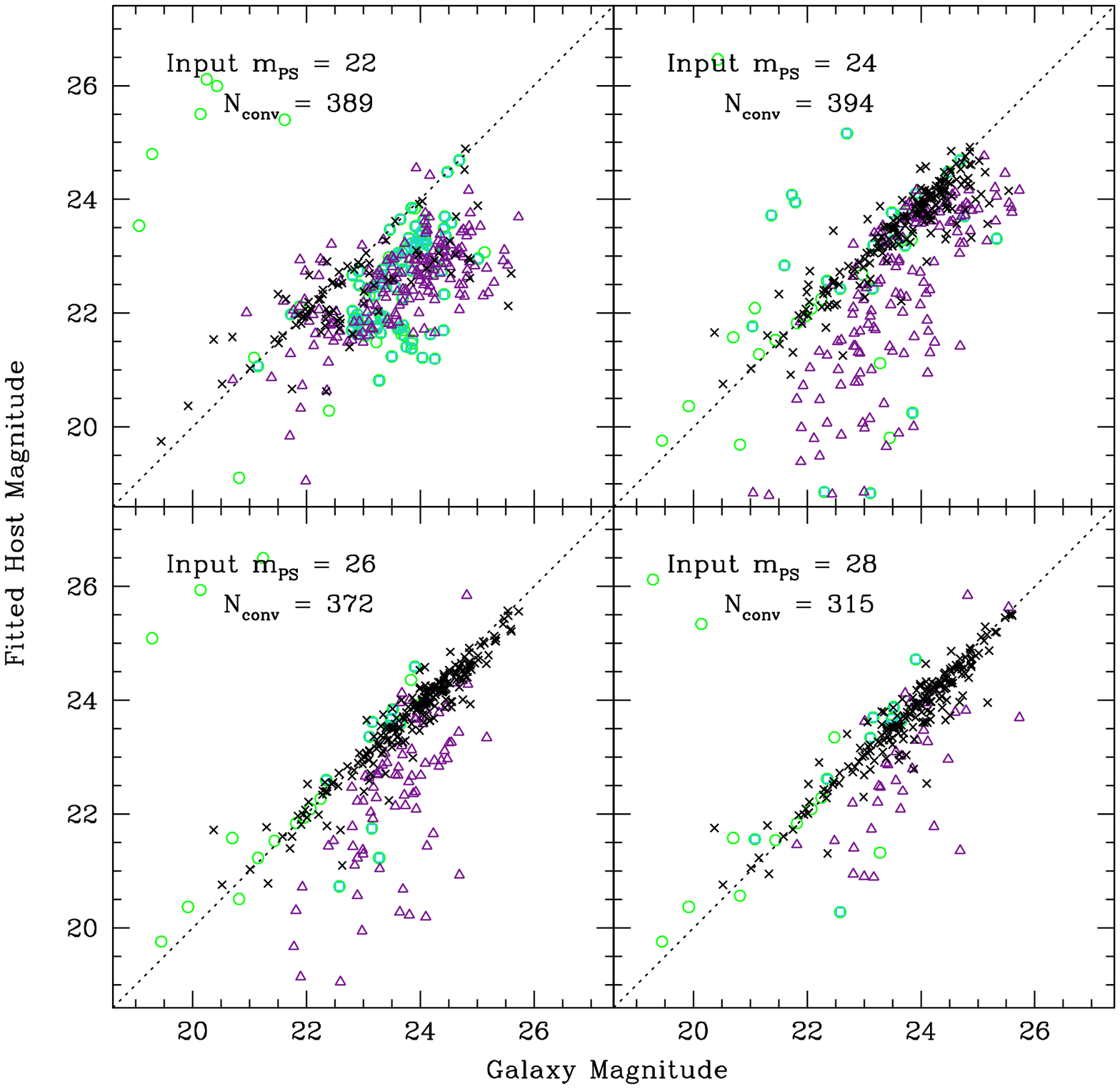}
\caption{
Fitted host galaxy magnitude versus actual galaxy magnitude for fits
to actual galaxies with added point sources. Excluding high $\mcsqnu$
values and effective radii with large fractional errors ensures a high
fraction of good fits, with moderate uncertainties. Green open circles
indicate fits with high $\mcsqnu$ values. Cyan open squares indicate
fits with excessive errors on the effective radius fit parameter
($\sigma_{r_e} \geq 0.8 \times r_e$). Purple open triangles represent
fits with both high $\mcsqnu$ and high $\sigma_{r_e}$. The dotted line
represents equal input and recovered magnitudes (i.e., perfect
recovery). The error cut removes some of the fits that do lie on the
1:1 line, regardless of the precise value for the $\sigma_{r_e}$ cut.
}
\label{m0mp_all} 
\end{figure}

Figure \ref{m0mp_all} shows the correlation between the fitted model's
host galaxy magnitude and the actual host magnitude (as determined
before the addition of the central point source). The percent
completeness of the low-error sample compared to the full sample
varies greatly across a range of contrast ratios between host galaxy
and central point source. For very faint central point
sources, the host galaxy magnitude is very well recovered, whereas for
very bright point sources the fraction of host galaxies with
well-recovered parameters is very low ($\sim 15$\%).

Our simulations show that the reduced \csqnu\ parameter is generally
not sufficient to distinguish good from poor host galaxy fits. If we
also limit the relative error of the effective radius parameter,
however, the fraction of selected galaxies that recover the true
magnitude of the host galaxy improves significantly. With the
combination of \csqnu \ and $\sigma_{r_e} / r_e$ parameters, we remove
more than 85\% of the fits with magnitude differences of 0.5 dex or
more from the actual galaxy magnitudes. The number of well-fitted
galaxies that are also removed from the sample ranges from 5\% to
20\%, depending primarily on how conservatively we choose our relative
$\sigma_{r_e}$ cut. Figure \ref{m0mp_all} reflects a cutoff value of
$\left( \sigr \right) = 0.8 $, which removes 88\% of poor fits and
12\% of the good fits from the $I_{\mbox{\small PS}} = 24$ sample.

\begin{figure}
\plotone{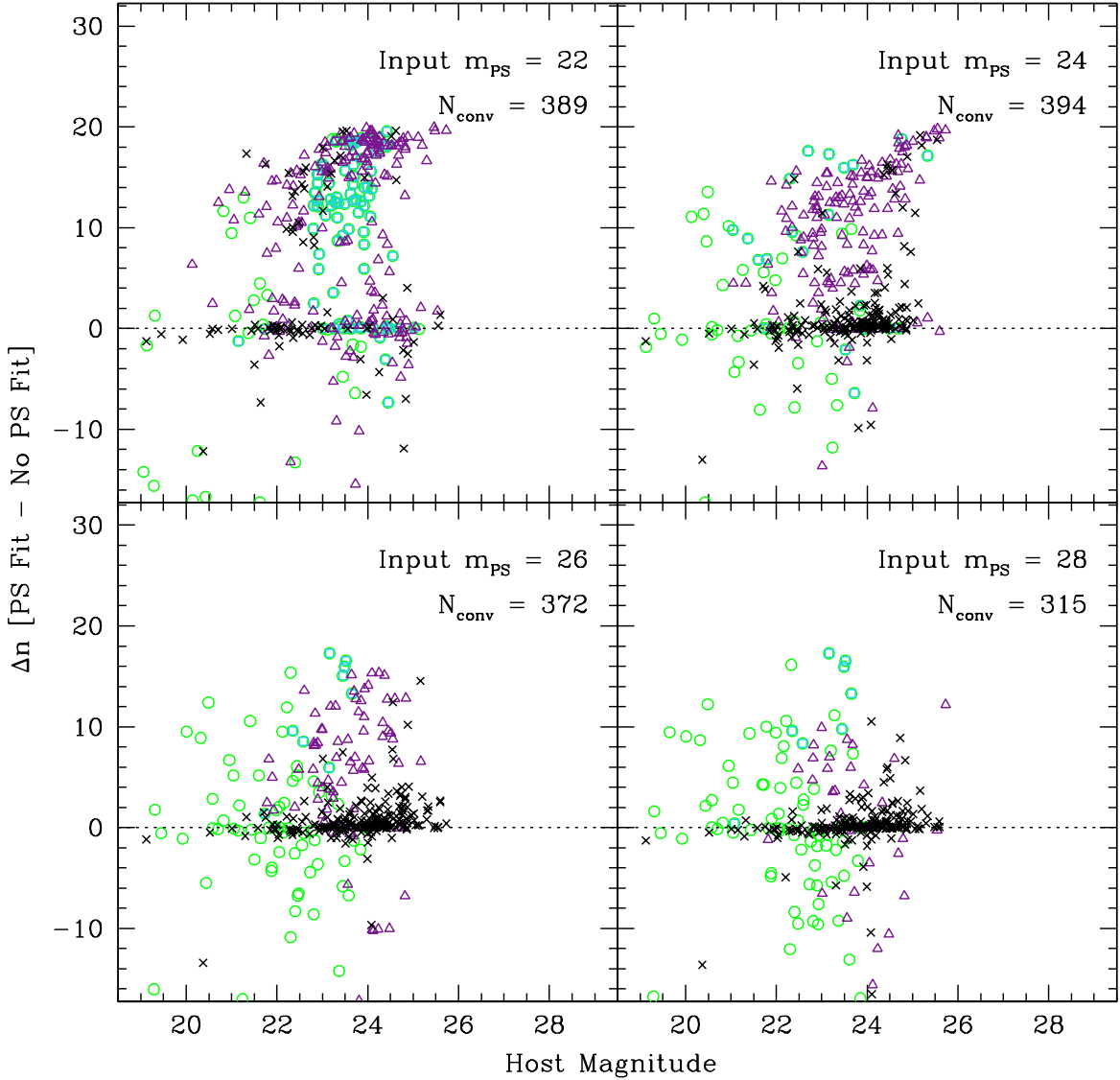}
\caption{Change in S\'ersic indices of host galaxies after adding
point sources with original host magnitude. Excluding high $\mcsqnu$
and $\sigr$ yields the correct S\'ersic index (a proxy for
morphological type). The color-coding of the points is the same as in
Figure \ref{m0mp_all}: green open circles for high $\mcsqnu$, cyan
open squares for high relative $\sigma_{r_e}$, and purple open
triangles for both high $\mcsqnu$ and relative $\sigma_{r_e}$. The
dotted lines at $\Delta n = 0$ indicate perfect recovery of the
S\'ersic index. As expected, the bright-point-source fits return
mostly unreliable host galaxy parameters, in contrast to the
faint-point-source fits, which return mostly reliable host galaxy
parameters. }
\label{dnpm}
\end{figure}

Figure \ref{dnpm} shows how well we recover the S\'ersic index,
$n$. The same error cuts (\csqnu\ and $\sigr$) efficiently reject the
inaccurate fits. The fraction of models whose fitted $n$ values
deviate by more than $\Delta n = 2$ compared to the $n$ values of the
unaltered galaxies increases significantly with increasing
point-source flux. That is, a very bright central point source can
easily be confused with a concentrated (high-$n$) S\'ersic
value.  For all values of the added point-source magnitude, however,
the same combination of \csqnu\ and relative $r_e$ error is sufficient
to remove 88\% of the inaccurate fits from the data, while removing
less than 15\% of the well-fitted models.

\begin{table}
\begin{center}
\begin{tabular}{|cc||cc|cccc|}\hline
Source $m_{PS}$& $N_{conv}$ & $\sigma_{m_{PS}}$ & $N_{1}$ & $\sigma_{m_{host}}$ & $\sigma_n$ & $\sigma_{r_e}$ & $N_{2}$ \\ 
\tiny{(AB)} & & & & & & \tiny{(pixels)} & \\ \hline
 21 & 374 &  0.33 & 210 &  1.09 &  2.10 & 15.6 &  45 \\
 22 & 389 &  0.48 & 273 &  0.12 &  0.97 & 21.8 &  85 \\
 23 & 397 &  0.27 & 359 &  0.15 &  0.67 & 11.0 & 138 \\
 24 & 394 &  0.15 & 328 &  0.20 &  0.54 & 26.7 & 184 \\
 25 & 382 &  0.17 & 328 &  0.14 &  0.40 & 15.9 & 223 \\
 26 & 372 &  0.27 & 295 &  0.17 &  0.44 & 8.2  & 215 \\
 27 & 355 &  0.47 & 271 &  0.12 &  0.27 & 7.9  & 215 \\
 28 & 315 &  0.87 & 228 &  0.10 &  0.22 & 6.3  & 192 \\
 29 & 286 &  1.20 & 193 &  0.09 &  0.27 & 1.6  & 161 \\ \hline
\end{tabular}
\caption{ RMS values for fit parameters vary with central
point-source magnitude. $N_{conv}$ is the number of fits which
converged (out of 450) for each input point-source magnitude
value. $N_1$ is the number remaining after applying a $\mcsqnu < 2.0$
cut; $N_2$ is the number remaining after further requiring $\sigr <
0.8$. The RMS value for the point-source magnitude is calculated from
$N_1$ galaxies, and the values for $\sigma_{m_{host}}$, $\sigma_n$,
and $\sigma_r$ are calculated from $N_2$ objects.  \emph{Note:}
Results for these magnitudes are based on a survey with the GOODS
depths \citep{giavalisco04}.}
\label{rms_params}
\end{center}
\end{table}

\begin{table}
\begin{center}
\begin{tabular}{|c||cccc|}\hline
Source $\lcontr$ & $\sigma_{m_{PS}}$ & $\sigma_{m_{host}}$ & $\sigma_n$ & $\sigma_{r_e}$ \\ 
 & & & & \tiny{(pixels)} \\ \hline
 -1.5 & 0.26 &  0.78 &  0.70 & 7.2 \\
 -1.0 & 0.31 &  0.29 &  0.38 & 6.2 \\
 -0.5 & 0.31 &  0.31 &  0.31 & 7.1 \\
  0.0 & 0.36 &  0.35 &  0.27 & 5.4 \\
  0.5 & 1.69 &  0.32 &  0.24 & 4.4 \\
  1.2 & 1.92 &  0.23 &  0.33 & 4.6  \\
  1.9 & 0.69 &  0.07 &  0.20 & 3.4  \\ \hline

\end{tabular}
\caption{ Determined RMS values for fit parameters to sources with
different $\lcontr$. }
\label{rms_params_CR2}
\end{center}
\end{table}

\subsubsection{Dependence of Parameters on Host:AGN Contrast Ratio}

\begin{figure}
\plotone{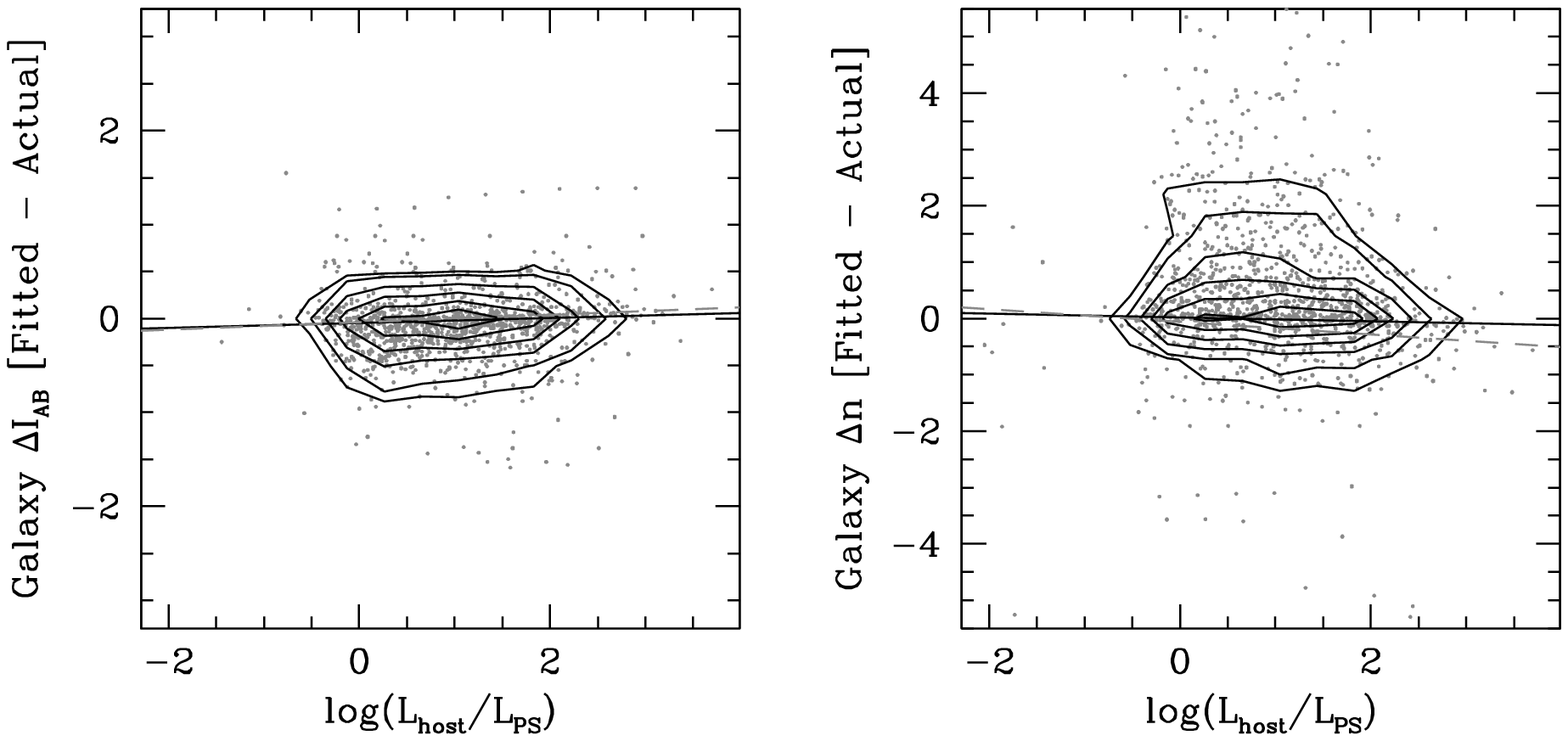}
\caption{ Accuracy of the host galaxy magnitude (left) and S\'ersic
index (right) as a function of the ratio of intrinsic galaxy
luminosity to the added point-source luminosity.  In both figures,
contour plots are overlaid on the data points. Weighted least-squares
fits to all data points (dashed lines) indicates that the recovery of
galaxy magnitude and S\'ersic index is very slightly correlated with
host-to-PS contrast ratio. Excluding points with $\lcontr < -0.6$, the
weighted least-squares fit (solid lines) is consistent with zero slope
(i.e., no correlation between Host-to-PS contrast ratio) for
both plots.  }
\label{contr_mn}
\end{figure}

Figure \ref{contr_mn} shows the relation of fitted host galaxy
parameters to the contrast ratio between the host galaxy and the added
central point source, for the 1491 (of 4050) sources that are not
eliminated by the combination of $\mcsqnu$ and $\sigr$ cuts. These
include galaxies with all possible values of input point-source magnitude.

The left panel of Figure \ref{contr_mn} shows the difference between
the fitted host galaxy magnitude and original galaxy magnitude for the
PS-added models. The recovered magnitudes are no different on average
than the correct parameters. For galaxies with contrast ratios of at
least $\lcontr <-0.6$, the host galaxy magnitude is well recovered,
such that a weighted least-squares fit to those data with contrast
ratios $\contr$ greater than 1:4 has a slope within 1\% of zero,
indicating no correlation between contrast ratio and recovered host
galaxy magnitude.  The scatter in the plot indicates that, while the
sample as a whole is well-recovered, individual galaxies may have
large differences in recovered parameters due to the presence of a
point source.  We see this result again in the right panel, which
shows the difference between fitted and original S\'ersic index as a
function of host-to-PS contrast ratio. The inaccuracy in S\'ersic
index is larger when the added central point source is more than 4
times as bright as the host galaxy: the slope of the weighted
least-squares fit is 6\% for $\Delta I_{AB}$ and 11\% for $\Delta
n$. With proper error cuts, and for galaxies brighter than $1/4$ of
the point-source magnitude, our host galaxy fits are not significantly
contaminated by the central point source. However, the presence of a
point source increases the uncertainty in fit parameters.

\begin{figure}
\plotone{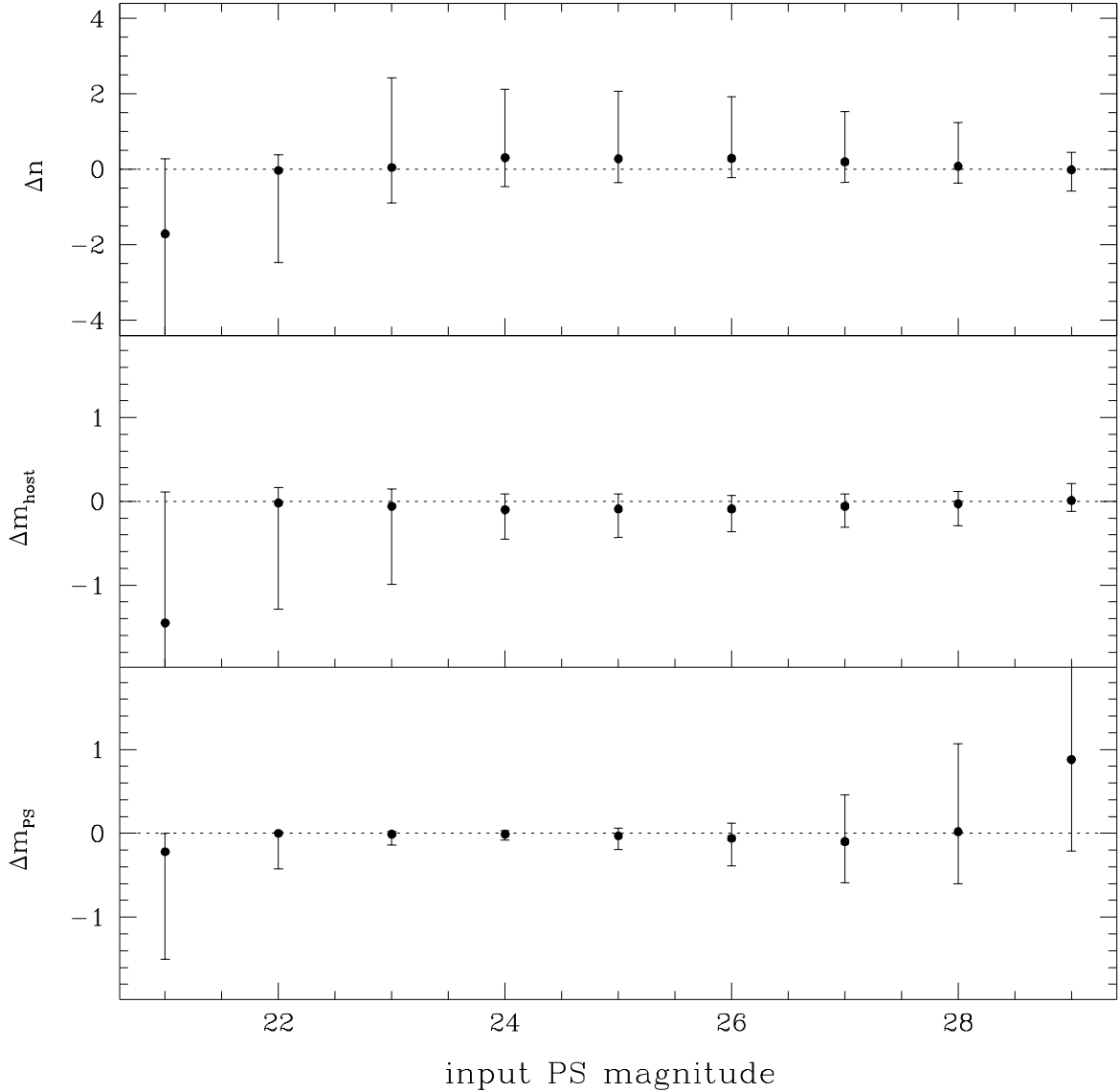}
\caption{Median change in S\'ersic index $n$ (upper), host magnitude
(middle), and point-source magnitude (lower) versus input point-source
magnitude. Each point represents the median value of all the galaxies
with the given input point-source magnitude whose fits pass the
\csqnu\ and $\sigr$ cut described in the text. The host S\'ersic
indices (top panel) are generally well-recovered for all but the
brightest input point sources. The host magnitudes (middle panel) are
also well-recovered in all but the brightest input point-source bin,
with the error bars (representing the width of the distribution
encompassing 68\% of sources) increasing with increasing point-source
strength. Point-source magnitudes (bottom panel) are recovered well
until the input magnitude is fainter than the published $10\sigma$
point-source detection limit of the data. }
\label{n_m_avg}
\end{figure}

Figure \ref{n_m_avg} shows the median change in fitted parameters with
input point-source magnitude. For very bright added point sources
$\left( I_{AB} = 21\mbox{, }\left<\mbox{Host:PS} \right> \approx
1:11\right)$, all fit parameters deviate from their
initially-determined values. The median S\'ersic index is low by
approximately $\Delta n = 1.75$, indicating that on average, galaxies
with such bright point sources could be classified as disk-dominated
even if they have more bulge-dominated intrinsic morphologies. The
distribution is also very wide in the brightest-point-source bin: for
such low host-to-PS contrast ratios, the S\'ersic index is essentially
completely uncertain. For fainter point-source magnitudes, recovery of
the S\'ersic index is far more reliable, although the uncertainty
(indicated by the width of the distribution in each bin) is
considerably larger than that reported by the fitting routine.

The ratio of host and point-source luminosities is also incorrect at
the brightest point-source magnitudes: in these situations the typical
fitted point-source magnitude is very close to the input value, and
the fitted host magnitude is too low (bright) by approximately 1.45
dex. Interestingly, the deviations in magnitude do not add to zero:
because bright point-sources are also slightly too bright, the total
magnitude is also underestimated (too bright) in the presence of a
very bright central point source.

For point sources brighter than $I_{AB} = 24$, the distribution of
recovered host galaxy magnitude is wide, such that the 68\% confidence
intervals span more than 1 dex in all bins. For those point sources
with magnitudes fainter than 24 (which is approximately equal to the
average magnitude of our initial galaxy sample), on the other hand,
the model's host galaxy magnitude is well recovered.

Since the faintest point-source magnitudes ($I_{AB} = 28$, $29$) are
below the $10\sigma$ point-source detection limit of our band, the
fitted point-source magnitude for these cases is often at least 1 mag
brighter than the input point-source magnitude, with 68\% confidence
intervals larger than 1 mag.

The S\'ersic indices for the host galaxies are within $\Delta n =
0.25$ on average for most values of the added point-source
magnitude. Since disks are generally more extended than bulges, the
S\'ersic parameter is lower (on average) when the central point source
is so bright that the central portion of the host galaxy is dwarfed by
the light in the wings of the central point source.  In all but the
brightest cases, however, the median difference between the model
S\'ersic index and the no-point-source index is smaller than the size
of the morphological bins discussed for a typical AGN sample
(e.g., \citet{sanchez04,ballo07}, S08), so it is unlikely that
a significant fraction has been mis-classified as disk-like or
bulge-like due to the presence of a central point source.

The rms values of the recovered S\'ersic indices and recovered host
magnitudes for each input point-source magnitude are given in Table
\ref{rms_params}.

\subsection{Fully Simulated Galaxies}

\subsubsection{Input Parameter Recovery}

For our 12,592 fully simulated GOODS galaxies, we apply the error cut
determined above, using a combination of \csqnu\ and $\sigr$ to
distinguish good fits from poor fits.  Within the single component
galaxies, this error cut results in 706 of 896 B type galaxies with
``good'' fits and 708 of 896 D type galaxies with ``good'' fits.

\begin{figure}
\plotone{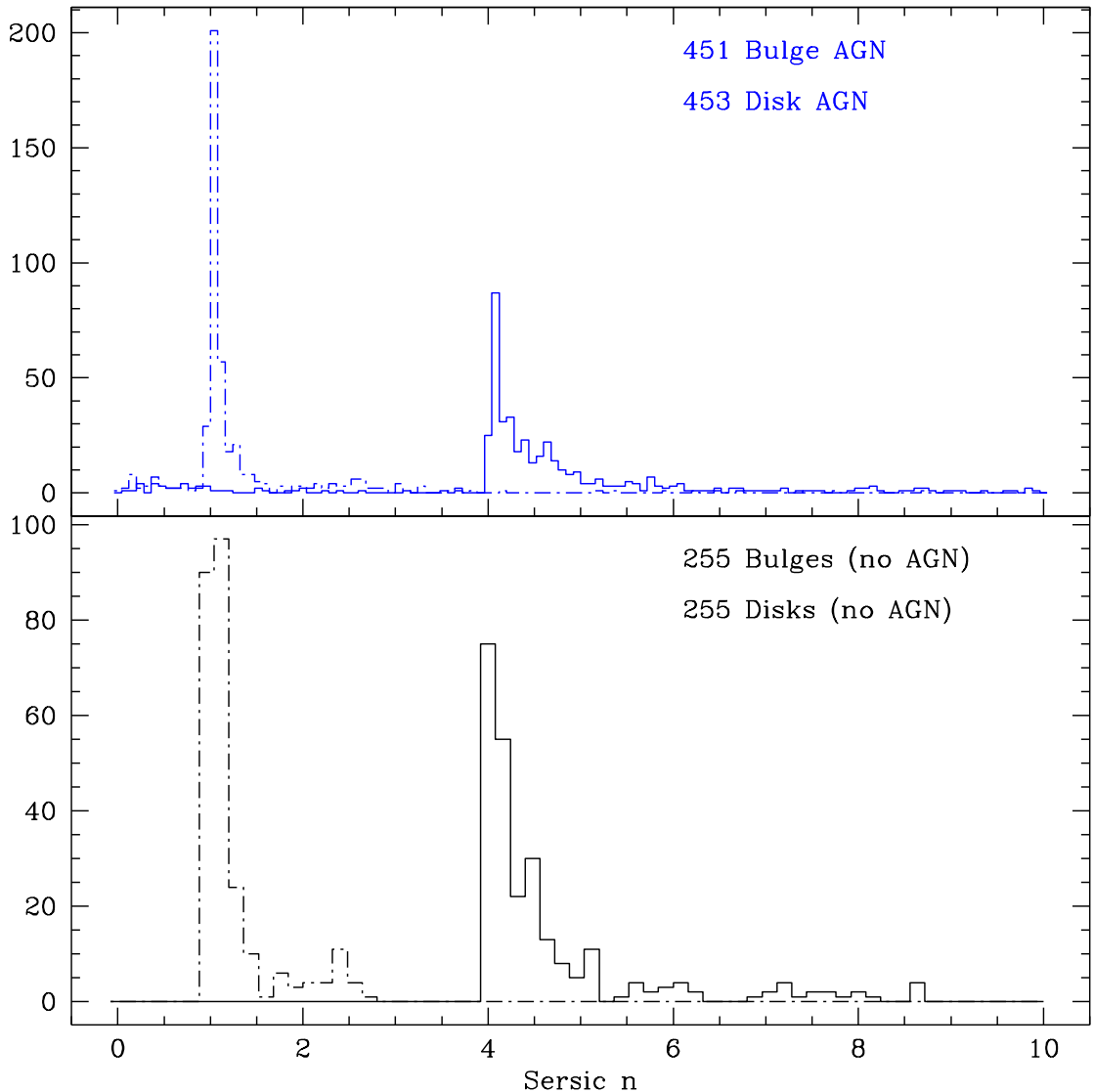}
\caption{Recovery of S\'ersic index $n$ for the single-S\'ersic fits
to simulated single-component B and D AGN host galaxies (blue, top)
and inactive galaxies (black, bottom). For the bulge-dominated B
galaxies (solid histograms), the average fitted S\'ersic index is
slightly higher than the input $n=4$ in both AGN and inactive
galaxies; however, these are unambiguously classified as
bulge-dominated galaxies, with fewer than 10\% of bulge host galaxies
mis-classified as disk-dominated galaxies and no bulge inactive
galaxies mis-classified as disks.  For the disk-dominated D galaxies
(dot-dash histogram), the recovery of the input $n=1$ is also
excellent, with 13\% of D host galaxies and 9\% of D inactive galaxies
having $n > 2.0$.  The separation between the typical fitted $n$ for B
and D galaxies is robust and indicates that we can use the S\'ersic
index as a strong indicator of galaxy morphology for single-component
local galaxies with and without central point sources.  }
\label{B_D_hist}
\end{figure}

The recovery of the S\'ersic index for both of these types of galaxy
models is shown for the $z=0.125$ sample in Figure \ref{B_D_hist}. If
we classify galaxies as bulge-dominated when their fitted $n > 2.0$
and disk-dominated when their fitted $n < 2.0$, the level of
misclassification based on these simulations is less than 15\% for AGN
host disk galaxies and less than 10\% for AGN host bulge
galaxies. Fewer than 10\% of inactive disks are misclassified as bulges
using the above cut; no inactive bulges are misclassified as
disks. For any local sample of single-component galaxies (the simplest
possible case), then, the S\'ersic parameter is a very reliable
indicator of morphological type.

\begin{figure}
\plotone{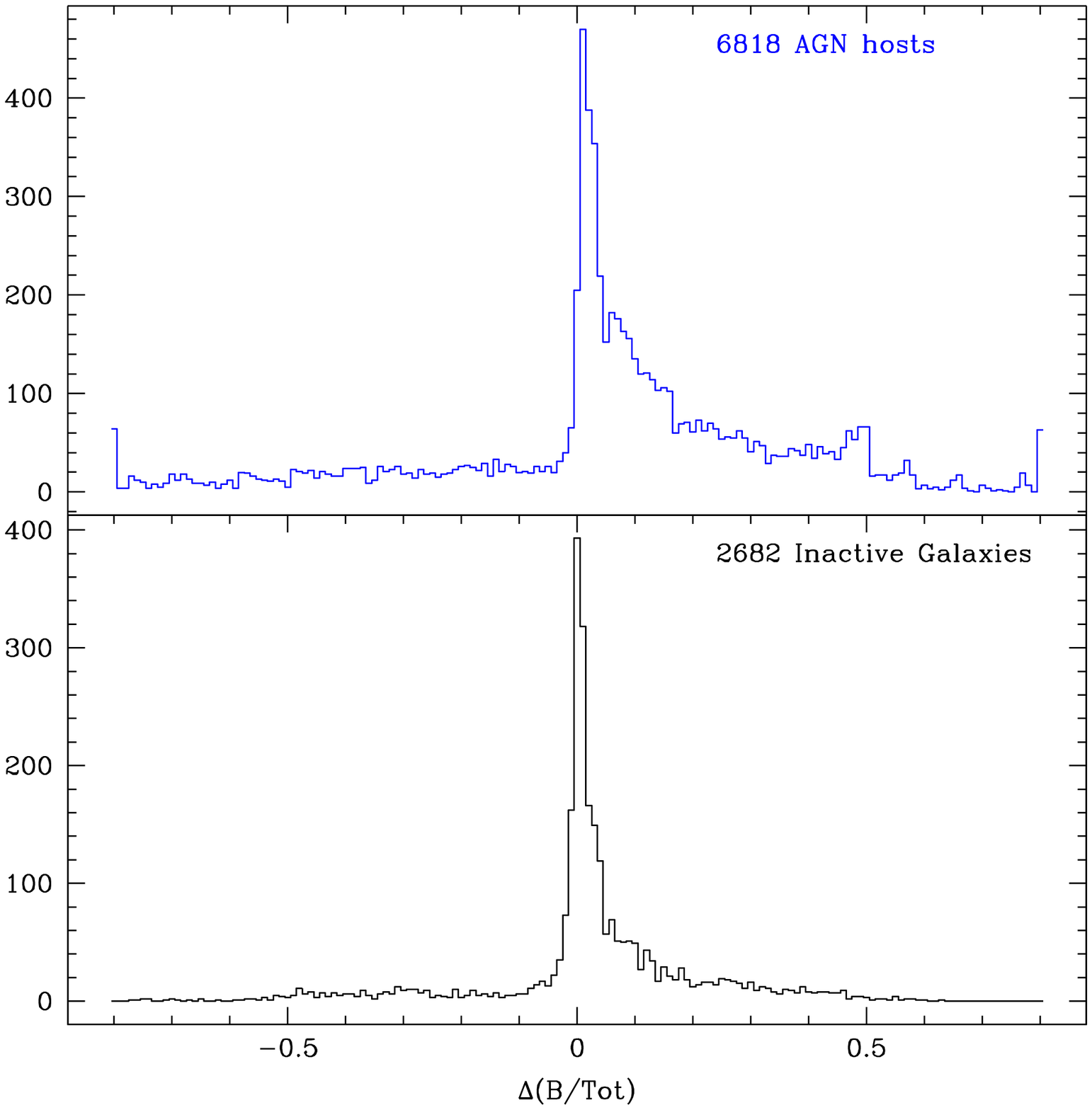}
\caption{ Recovery of input \BT\ ratio for the fully-simulated
two-component galaxies. We define $\Delta \mBTlet > 0$ to mean the
fit was more bulge-like than the input galaxy. A Gaussian fit to the
Host galaxy distribution gives $\sigma_{BT} = 0.04$, but there is a
large asymmetric tail on each distribution indicating that the samples
are skewed toward more bulge-like fits than the intrinsic
distribution. 36\% of host galaxies and 21\% of inactive galaxies have
$\Delta \mBTlet > 0.15$.}
\label{BDrat_hist}
\end{figure}

The recovery of the $z=0.125$ sample of two-component (B+D) galaxies
is shown in Figure \ref{BDrat_hist}. The fitted \BT\ light ratio,
B/Tot $= L_B/\left(L_B + L_D\right)$, is within $\left| \Delta \mBTlet
\right| \leq 0.15$ for 50\% of our AGN host sample and 66\% of our
inactive sample. A Gaussian fit to the data in Figure \ref{BDrat_hist}
gives an estimate of uncertainty in the \BT\ parameter of
$\sigma_{\mbox{\tiny B/Tot}} = 0.04$. However, the contamination from
very deviant fits (i.e., the wings of the histogram in Figure
\ref{BDrat_hist}) is large. Approximately 14\% and 13\% of host and
inactive galaxies remaining in the sample after our error cut have
$\Delta \mBTlet < -0.15$, and 36\% of host galaxies and 21\% of
inactive galaxies have $\Delta \mBTlet > 0.15$.

\begin{figure}
\plotone{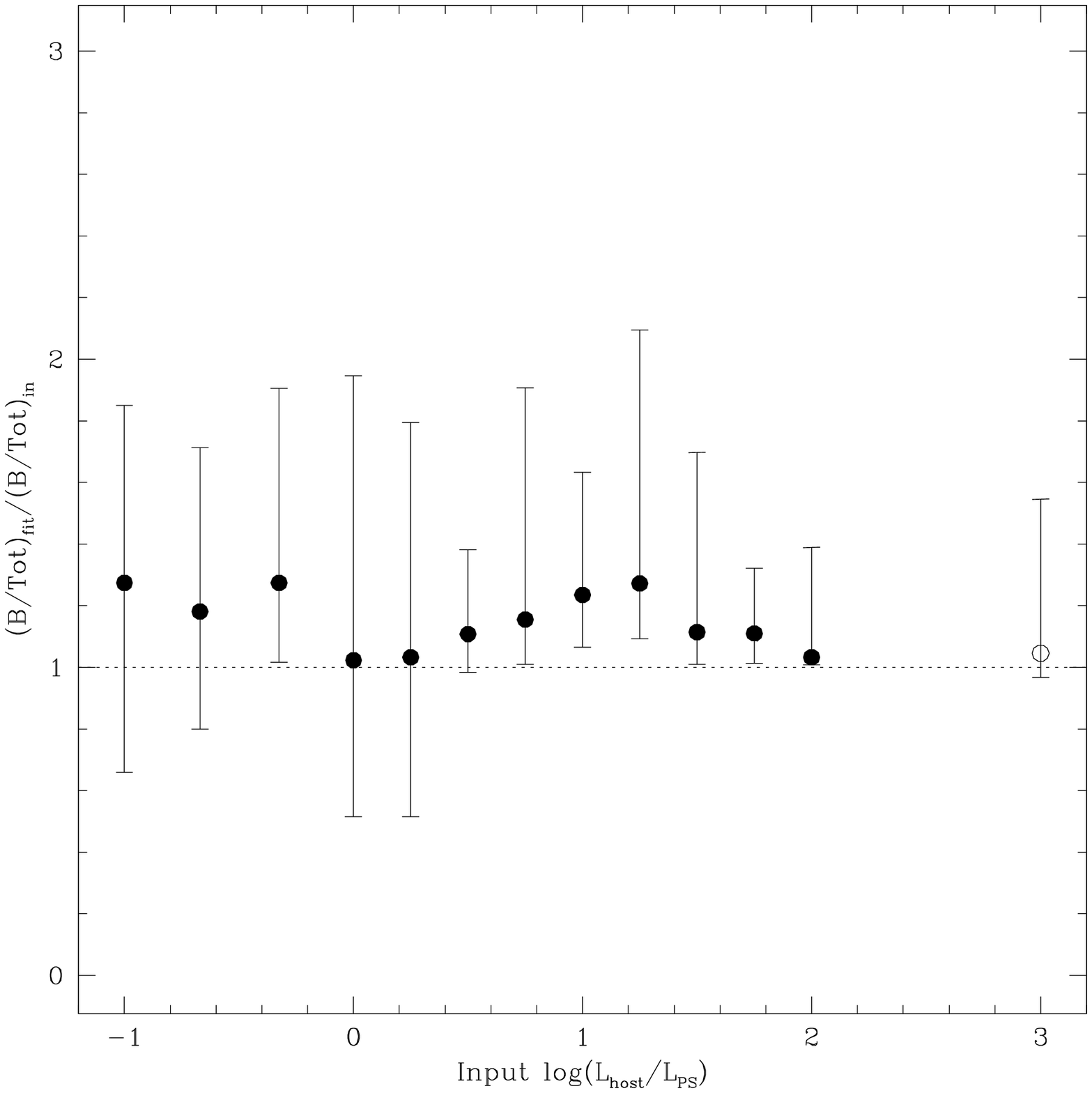}
\caption{Ratio of fitted-to-input \BT\ vs. input host-to-PS contrast
ratio.  Points plotted are the median values in each bin, with error
bars marking the range of values containing the central 68\% of the
points in each bin. The open circle shows simulated galaxies with no
added point source; this bin recovers input \BT\ ratio closely, with a
tail of high \BT\ values. For simulated host galaxies, the recovered
\BT\ ratio is generally high by 10-20\%, and the distribution of
points in each bin is wider for galaxies where the central point
source is at least as bright as the host galaxy.  }
\label{fBT_CR}
\end{figure}

Figure \ref{fBT_CR} shows the ratio of fitted-to-input \BT\ flux ratio
as a function of the intrinsic contrast ratio between the host galaxy
and the central point source. The median and 1$\sigma$ error bars
(defined here as the values encompassing 68\% of the points in each
bin) for the distribution of simulated galaxies without central point
sources are plotted for comparison at a value of $L_{host}/L_{PS} =
3$. In general, the presence of a central point source skews the fits
to slightly high values of the \BT\ ratio. For galaxies where the
central point source is very bright, the spread is higher, such that a
galaxy may fit to a \BT\ ratio twice the intrinsic value and still be
within 1$\sigma$ of the median value. Based on this, one could attempt
to correct the \BT\ ratio for the presence of a central point source,
but a \BT\ ratio fitted to a host galaxy could be overestimated by as
much as 100\%.

This large tail of high fitted \BT\ values has several causes. The
radius of the bulge may converge to an unphysically high value; this
causes the luminosity \BT\ ratio to skew high.  This appears to be
slightly more common when the intrinsic luminosity of the galaxy is
disk-dominated, i.e., with $\mBTlet \leq 0.3$. In addition, if
the total source magnitude is not well-conserved during the fit, the
\BT\ ratio typically deviates from the input value by at least 10\%
(high or low). This is more likely to occur when the bulge and disk
components have approximately equal luminosity, or the bulge is slightly
brighter [$0.45 \leq \mBTlet \leq 0.7$].

About 18\% of the two-component AGN host fits result in either a bulge
or disk half-light radius that is unphysically low given the input
sizes of both in our sample. These fits span the entire range of
$\Delta \mBTlet$ values, and are far more common when the central
point source is intrinsically bright. For the locally-defined sample,
79\% of B+D bulge components and 63\% of B+D disk components with
fitted $r_e < 5.0$ pixels are those with the brightest central point
source; this comprises 26\% of all two-component sources with the
brightest central point source. It is worth noting that these fits
typically show residuals that clearly indicate a poor fit, and they
can therefore be removed from the well-fitted sample on visual
inspection, or followed up with individual fitting to improve the fit.

\begin{figure}
\plotone{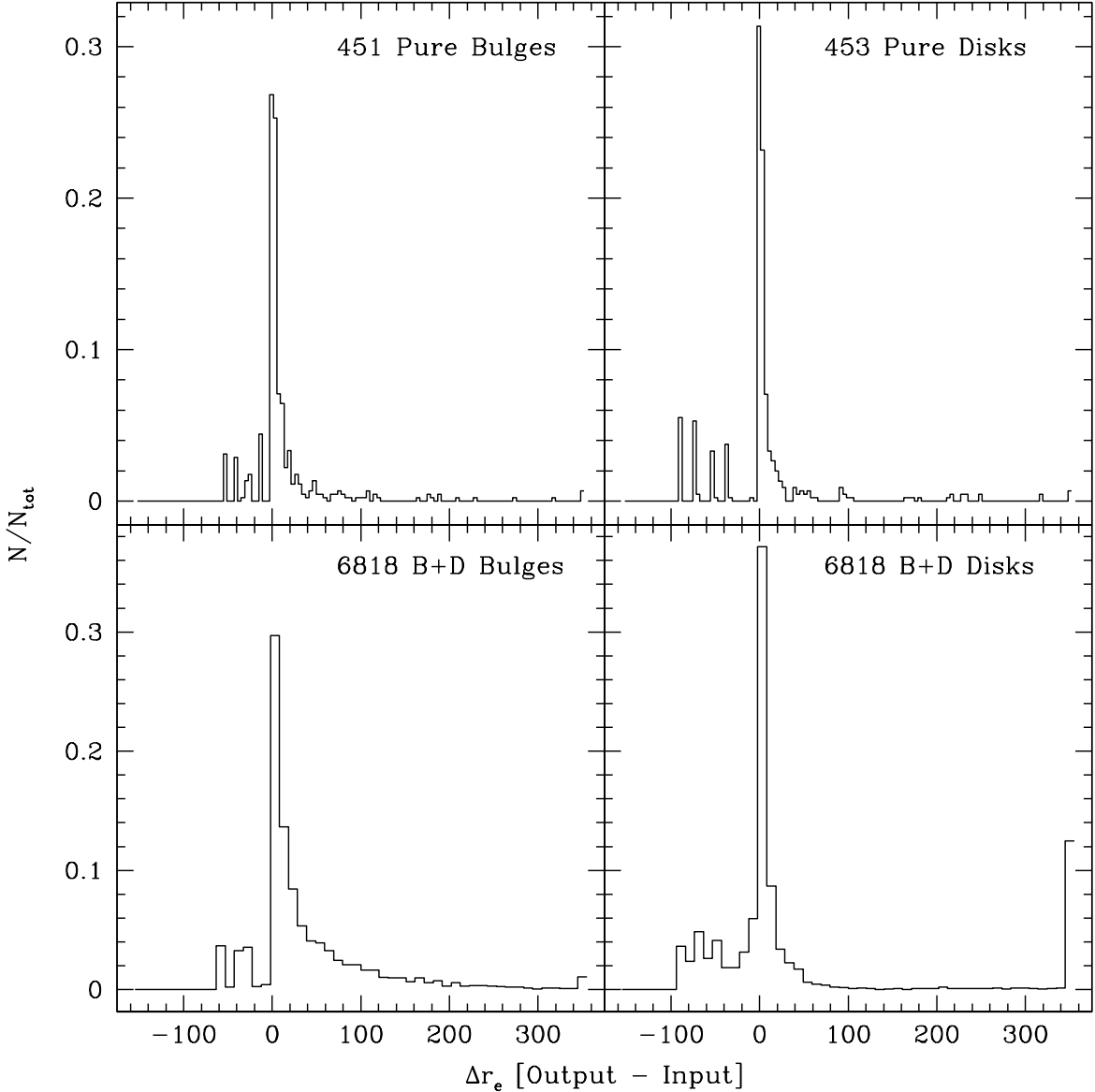}
\caption{ Histograms of the change in half-light radius for pure bulge
(top left), pure disk (top right), B+D bulge (bottom left) and B+D
disk (bottom right) fits to simulated AGN hosts. The majority of fits
converge to a value that is equal to or greater than the input value
of the half-light radius for that source or source component. The
small discrete peaks at negative $\Delta r_e$ values are comprised
mostly of fits for which $r_e$ converged to an unphysical value for
our sample, $r_e < 5$ pixels. Single-S\'ersic fits typically recover
the sizes of pure bulges and pure disks with about the same level of
success. However, the local sample of two-component galaxies indicates
that bulge component radii are more likely to converge to a value that
is at least 50 pixels too high, or at least twice the intrinsic
radius. Local disk components are more likely to be underestimated,
with a lower $\sigma$ for the high-$\Delta r_e$ distribution than for
B+D bulges. }
\label{rdiff_hist}
\end{figure}

The recovery of bulge and disk sizes in our sample is shown in Figure
\ref{rdiff_hist}. The distribution of $\Delta r_e$ for
single-component galaxies (pure bulges and disks) is sharply peaked at
a value of $\Delta r \approx 0$ (in pixels). A total of 66\% of pure
bulges and 65\% of pure disks recover their input radii within $-1
\leq \Delta r_e \leq 12$ pixels. The discrete peaks at negative
$\Delta r_e$ represent those fits that converged to unphysically low
values ($r_e < 5.0$ pixels for the local sample). Less than 5\% of
single-component fits converged to extremely large radii ($\Delta r_e
> 150$ pixels).

The radius parameter converges to extremely high values far more
frequently for the two-component fits, and with higher frequency for
the bulge components of two-component galaxies (20\%, B+D bulges) than
for the disk components of two-component galaxies (13\%, B+D
disks). The high-$\Delta r_e$ tail for B+D bulges is a result of
confusion with the central point source of the simulated AGN host. Not
only are bulges more compact than disks within our B+D galaxies (and
typically within real galaxies as well), but the slope of their light
profile more closely resembles the light profile of the ACS PSF than
does an exponential disk profile. This can lead to uncertain bulge
radii and magnitudes, especially when the intrinsic luminosity of the
bulge is similar to that of the central point source. Disk components
of B+D galaxies are not as readily confused with the central point
source: their fit typically fails when the surface brightness of the
disk component is too low compared to the bulge and point source
components.

\begin{figure}
\plotone{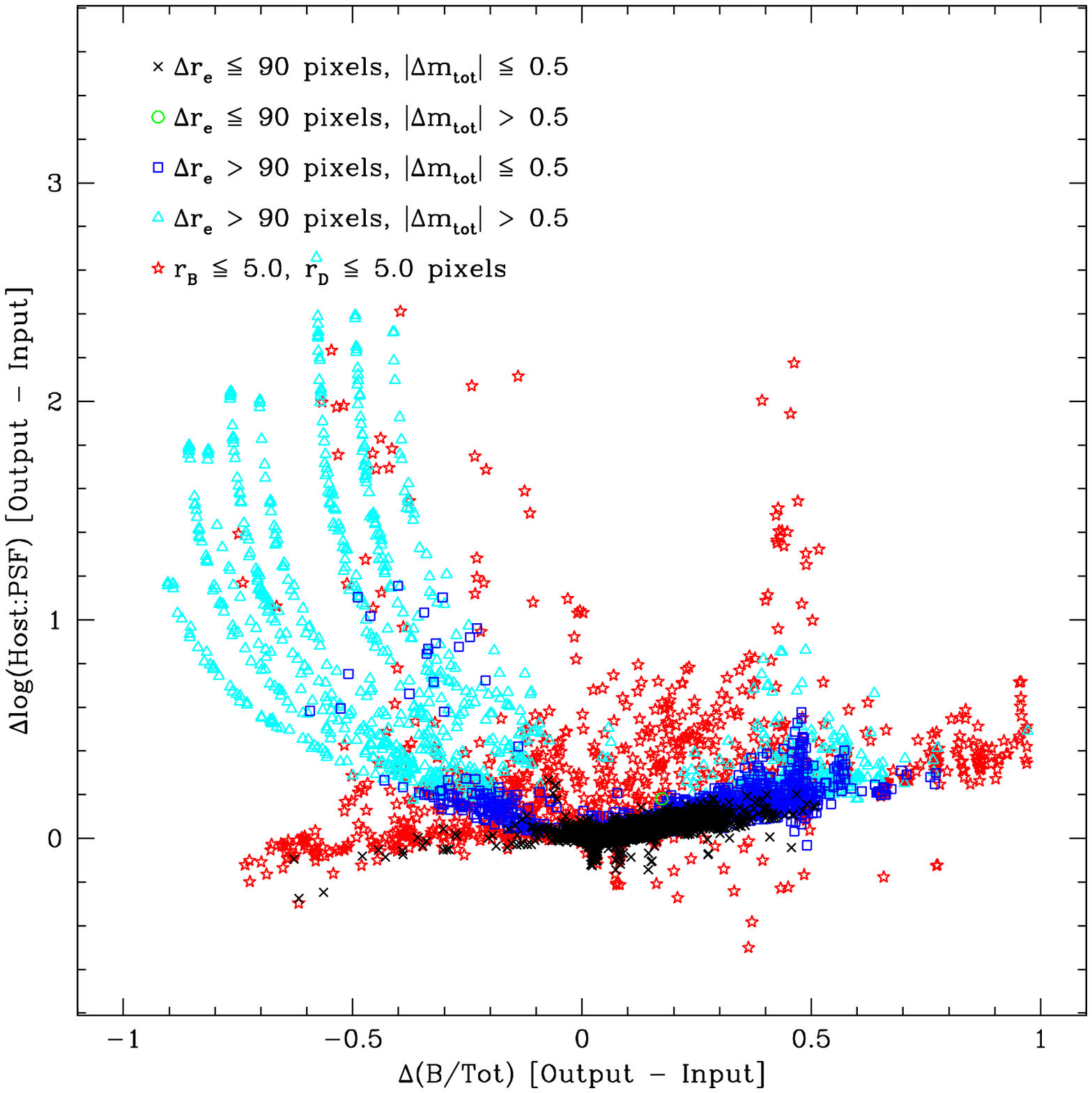}
\caption{ Deviation from input host-to-PS contrast ratio versus
deviation in \BT\ ratio. Fits located at (0, 0) perfectly recovered
the input \BT\ ratio and the Galaxy-AGN luminosity ratio
$\contr$. Requiring that the difference between the half-light radii
of the host be no more than 90 pixels larger than that observed for
the entire source ($\Delta r_e \leq 90$), that the total source
magnitude luminosity be conserved to better than $\left|\Delta
m_{tot}\right| \leq 0.5$ mag, and that the radii of neither the bulge
nor the disk component be less than 5 pixels (typically indicative of
a poor fit) cuts 81\% of sources with $\left|\Delta
\left(\mbox{B}/\mbox{Tot}\right)\right| > 0.15$, and retains 83\% of
sources with $\left|\Delta \left(\mbox{B}/\mbox{Tot}\right)\right|
\leq 0.15$.}
\label{dCR_dBT}
\end{figure}

Figure \ref{dCR_dBT} demonstrates a possible method of separating good
from poor two-component fits. After removing those fits where either
the bulge or disk converged to $r_e < 5.0$ pixels, additionally
requiring that the total source magnitude be conserved to within
$\left|\Delta m_{tot}\right| < 0.5$ (determined by comparing the
fitted total source magnitude with a measured, aperture-corrected
magnitude), and/or that the flux-weighted effective radius of the fit
be no more than 90 pixels larger than the measured half-light source
radius\footnote{This value was chosen because it is twice the observed
radius of the largest source in our sample.}  removes 81\% of sources
with $\left|\Delta \mBTlet\right| > 0.15$ and retains 83\% of sources
within 0.15 of the input \BT\ ratio. This also constrains the fitted
host-to-PS contrast ratio to within 8\% of the input value for 95\% of
the remaining ``good'' sample. However, this selection biases the
remaining sample toward sources with faint central nuclei and
preferentially retains bulge-dominated sources with intrinsic $\mBTlet
> 0.7$. In general, the result of two-component fits should be used
with caution, even for a local, well-understood sample.

\subsubsection{Detection of Central Point Sources}

\begin{figure}
\plotone{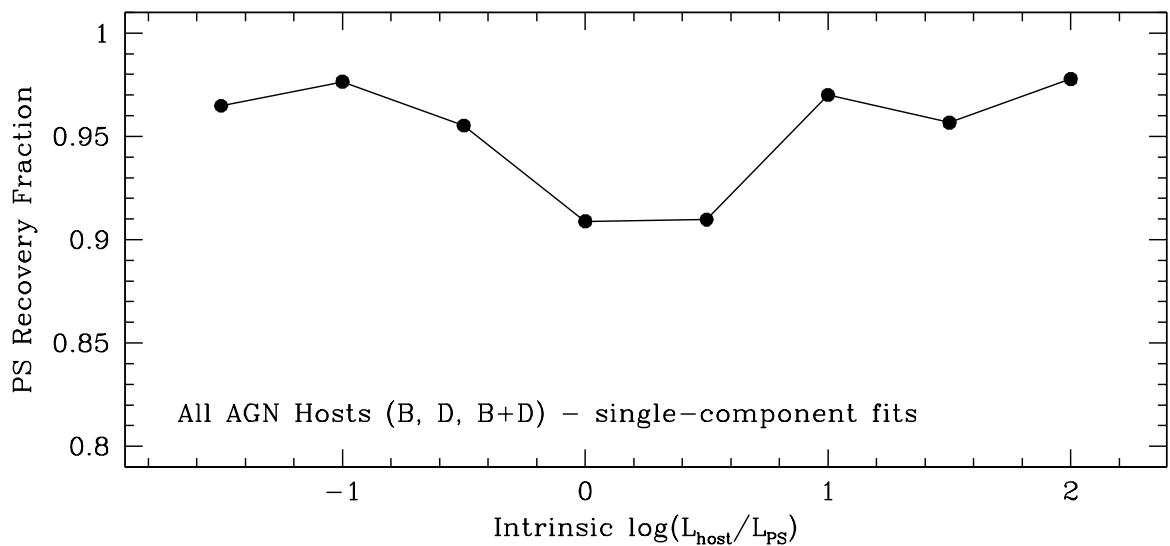}
\caption{ Fraction of point sources recovered for single-component
fits to all simulated AGN host galaxies. The fraction of recovered
point sources is slightly lower when the with input $\lcontr$ is near
zero ($L_{PS} \approx L_{host}$ or slightly fainter). However, even in
these cases, the fits detect at least 90\% of point sources.
}
\label{PSfrac_all}
\end{figure}

We established in \S 4.1.1 that central point sources can be recovered
on average for a sample of simulated host galaxies created from the
addition of a central point source to a real GOODS inactive galaxy,
although the fraction of recovered point sources varies depending on the
luminosity of the added point source. Using the \csqnu\ cut previously
determined from this sample, we now examine the recovery of point
sources with redshift and intrinsic morphological type. Figure
\ref{PSfrac_all} shows the fraction of fits in which the central point
source was detected for our entire sample of fully simulated
galaxies. In this case, our criteria for a ``detected'' point source
is simple: the point source is considered undetected only if the
magnitude parameter converged to a value considerably fainter than the
detection limit of our simulated data ($m_{PS} = 27.1$ for our sample;
\citet{giavalisco04}). Thus Figure \ref{PSfrac_all} shows the fraction
of fits for which any point-source magnitude was recovered, rather
than those fits for which the magnitude is close to the intrinsic
magnitude.

Figure \ref{PSfrac_all} indicates that the point source is detected
within the fit at least 90\% of the time for all values of input
$\lcontr$. The fraction is lowest when the point source and host
galaxy have about equal luminosity. The detection fraction increases
to more than 95\% at the highest values of input $\lcontr$, even for
cases in which the point source is 100 times fainter than its host
galaxy.  This is due to the fact that the convergence fraction is
lower for these bins; fits where the central point source may not have
been detected are less likely to converge, driving the detection
fraction higher for those fits that did converge.

\begin{figure}
\plotone{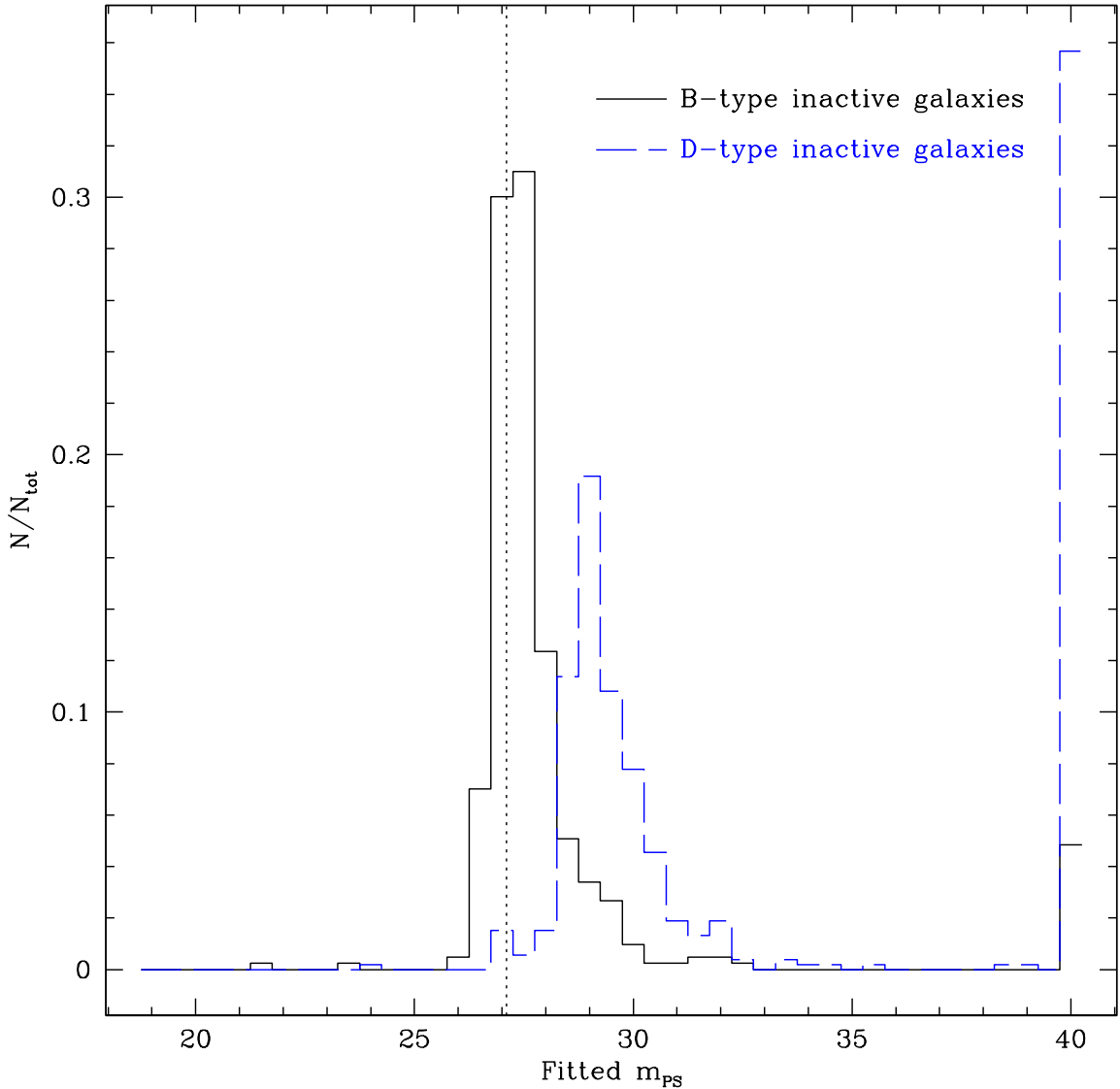}
\caption{ Distribution of fitted point-source magnitude for simulated
\emph{inactive} galaxies. Only 1\% of pure disk inactive galaxies
(blue dashed histogram) recovers point-source magnitudes brighter than the
$10\sigma$ detection limit (dotted line) for the GOODS survey (from
which our noise properties are taken). When the galaxy is a pure bulge
(black histogram), however, a point source is found above the
$10\sigma$ threshold in approximately 25\% of fits. The overall
percentage of fits where a point-source is found when none exists is
12\% (for both pure bulges and pure disks combined), but this number
is strongly dependent on intrinsic morphology.  }
\label{PSp0_hist}
\end{figure}

While we are primarily interested in the recovery of host galaxy
parameters and optical detection of central AGNs, we also want to
characterize the fraction of spurious optical point-source
identifications. Thus, Figure \ref{PSp0_hist} shows the results of a
S\'ersic + PS fit to \emph{inactive} galaxies with no intrinsic
central point source. We find a significant difference in the number
of false point-source detections between pure disk and pure bulge
galaxies. Only $\sim 1$\% of pure disk inactive galaxies
detect\footnote{We define ``detected'' as having been fit with a
magnitude brighter than the published $10\sigma$ point-source
detection limit of the GOODS survey (from which we take our noise
properties).} a central point source when none are present. However, we
find a point source nearly one-quarter of the time when the simulated
galaxy is a pure bulge. Because our total sample is composed of equal
numbers of pure bulge and pure disk galaxies, this gives an overall
false detection of a central point source in 12\% of inactive
galaxies. However, this effect is strongly dependent on morphology:
fits to elliptical galaxies are far more likely to detect a faint
point source when none are present than fits to disk galaxies. Host
galaxy studies estimating the uncertainty in the fraction of hosts
with detected optical point sources should therefore consider the
morphological composition of their data sample: if the sample contains
mostly elliptical galaxies, the number of false detections could be as
high as 25\%.

\subsubsection{Redshift Effects on AGN Host Galaxies}

At higher redshifts, surface-brightness dimming causes loss of
extended components, and an unresolved central AGN corresponds to a
larger fraction of a galaxy's central region at high redshift for
fixed spatial resolution. Because our GOODS data sample (S08), as well
as those of others \citep{sanchez04,grogin05,ballo07,alonsoherrero08},
extends to $z \approx 1$, we consider the quantitative effects of
redshift on morphological classification.

\begin{figure}
\plotone{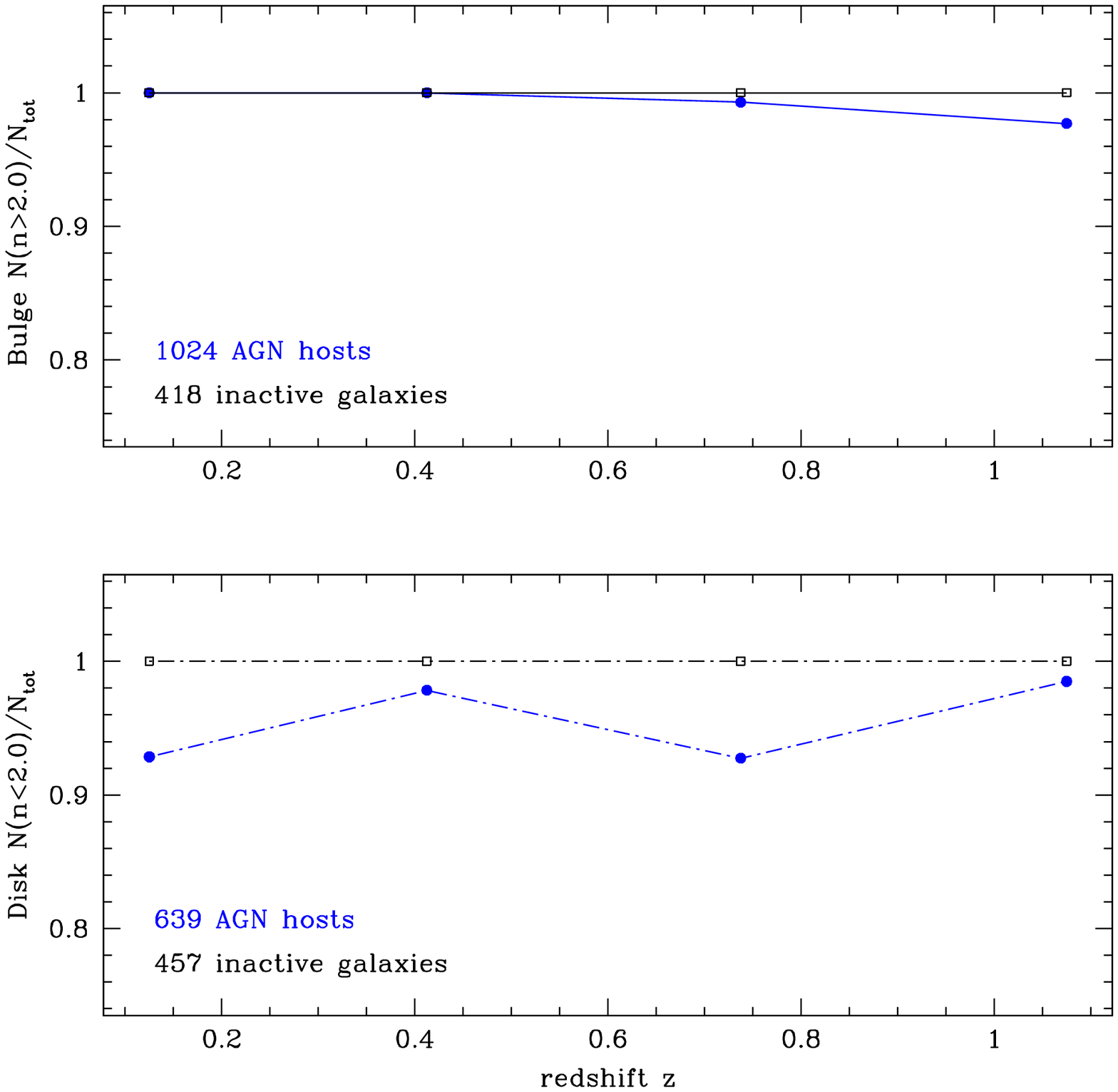}
\caption{ 
Dependence of galaxy classification on redshift for
simulated bulge galaxies ($n_{\mbox{\tiny in}}=4$; top, solid lines)
and disk galaxies ($n_{\mbox{\tiny in}}=1$; bottom, dot-dashed lines)
with (blue) and without (black) central point sources. The y-axis is
the fraction of galaxies correctly identified as bulges or disks given
a threshold of $n>2.0$ for bulge-dominated classification and $n<2.0$
for a galaxy with a significant disk. Although we have here adopted a
cutoff of $n=2.0$, the results change by less than 5\% when we change
the cutoff to values between $1.5 < n < 3$.
}
\label{nvz_B_D}
\end{figure}

After applying data cuts requiring $\Delta r_e \leq 90$ pixels and
$\left|\Delta m_{tot}\right| \leq 0.5$ to our results at all
redshifts, we assess the reliability of using the S\'ersic index to
classify single-component galaxies at different redshifts in Figure
\ref{nvz_B_D}. We have ``classified'' a galaxy as bulge-dominated if
it has $n > 2.0$ and disk-dominated if it has $n < 2.0$. Figure
\ref{nvz_B_D} then shows the fraction of correctly classified B- and
D-type galaxies at each redshift. In general, the classification of
pure bulges and pure disks is highly reliable for AGN host galaxies
and point sources. At all redshifts, disks and bulges are correctly
classified at least 90\% of the time using these data cuts. Without
these cuts, the misidentification rate increases to as much as 15\%.

\begin{figure}
\plotone{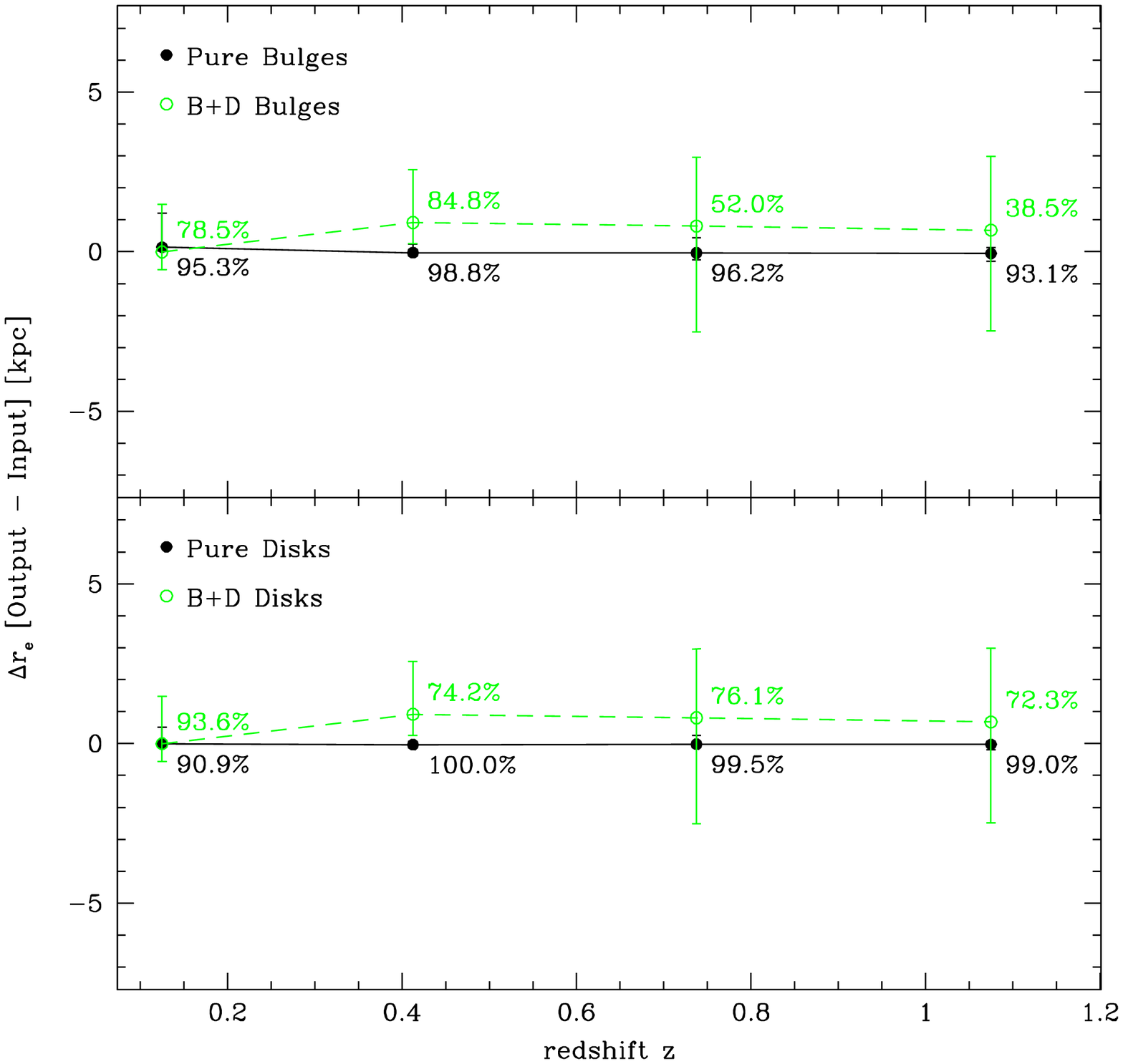}
\caption{ Recovery of input radii for pure and B+D bulges (top panel)
and disks (bottom panel). Each point represents the peak of the
distribution of single-component (black, solid lines) and
two-component (green, dashed lines) $\Delta r_e$ at each redshift. The
numbers next to each point indicate the percentage of completed fits
with $r_e > 5$ and $\Delta r_e \leq 90$ pixels, so that the fit
converged to a physical value and would not be rejected by eye based
on the fit residuals. While median fitted bulge and disk sizes are
generally accurate to within 1 kpc at all redshifts, the presence of a
central point source increases the spread in the distribution by up to
a factor of 3. In addition, the majority of bulge fits within B+D
galaxies at $z \approx 1$ are not able to converge to a sensible value
for the bulge radius.}
\label{rpeak_zavg}
\end{figure}

The average recovery of input half-light radius for all our simulated
galaxies (pure bulges and disks as well as B+D bulges and disks) is
shown in Figure \ref{rpeak_zavg}. Each point represents the peak of
the distribution of $\Delta r_e$ values for a single galaxy
type/component in each redshift bin. The error bars give the limits of
the central 68\% of data points; the peak of each distribution is
generally not coincident with the numerical ``center'' of each
bin. Single-component S\'ersic fits to pure bulges and disks recover
the radius reliably at all redshifts. Fitted radii to bulge and
disk components in B+D hosts recover the intrinsic radius of the
component to within 1 kpc on average; however, the errors are larger
and these fits are more uncertain. For example, the width (containing
68\% of sources) of the $\Delta r_e$ distribution of $z \approx 1$
pure disks is less than 1 kpc, whereas the width of the $z \approx 1$
distribution of $\Delta r_e$ for disk components in B+D hosts is
approximately 3 kpc. Although the average recovery of bulge and disk
component radii in B+D galaxies is accurate to within 1 kpc,
individual radii are highly uncertain.
 
The percentage of sources remaining after removing fits with incorrect
total magnitudes and unphysically low or high values of $r_e$, as
previously discussed, is also shown for each bin in Figure
\ref{rpeak_zavg}. The number of fits for which the fitted $r_e < 5$
pixels remains relatively constant with redshift, but the number of
fits for which the fitted $r_e$ value diverges increases with
redshift, more dramatically so for bulges than for disks. In the $z
\approx 1$ sample, fewer than 40\% of bulges within two-component fits
have fitted radii within a reasonable physical range. This is likely
due not just to the presence of the additional disk component, but
also to the fact that the bulge is much smaller at higher redshift
compared to the size of the central point source. In these cases, the
bulge and the point source can be confused, often resulting in a
divergent bulge fit. This happens less frequently for disks for two
reasons: first, disks are generally more extended than bulges, and
second, the radial profile of the ACS PSF more closely resembles that
of a de Vaucouleurs bulge than an exponential disk.

Because different fractions of galaxies are removed from the sample at
different redshifts (and for different bulge sizes), the morphological
composition of a galaxy sample may be altered from its intrinsic
composition simply by applying a reasonable cut on the morphological
fit parameters, even if this cut is applied evenly (as it is in Figure
\ref{rpeak_zavg}).

\begin{figure}
\plotone{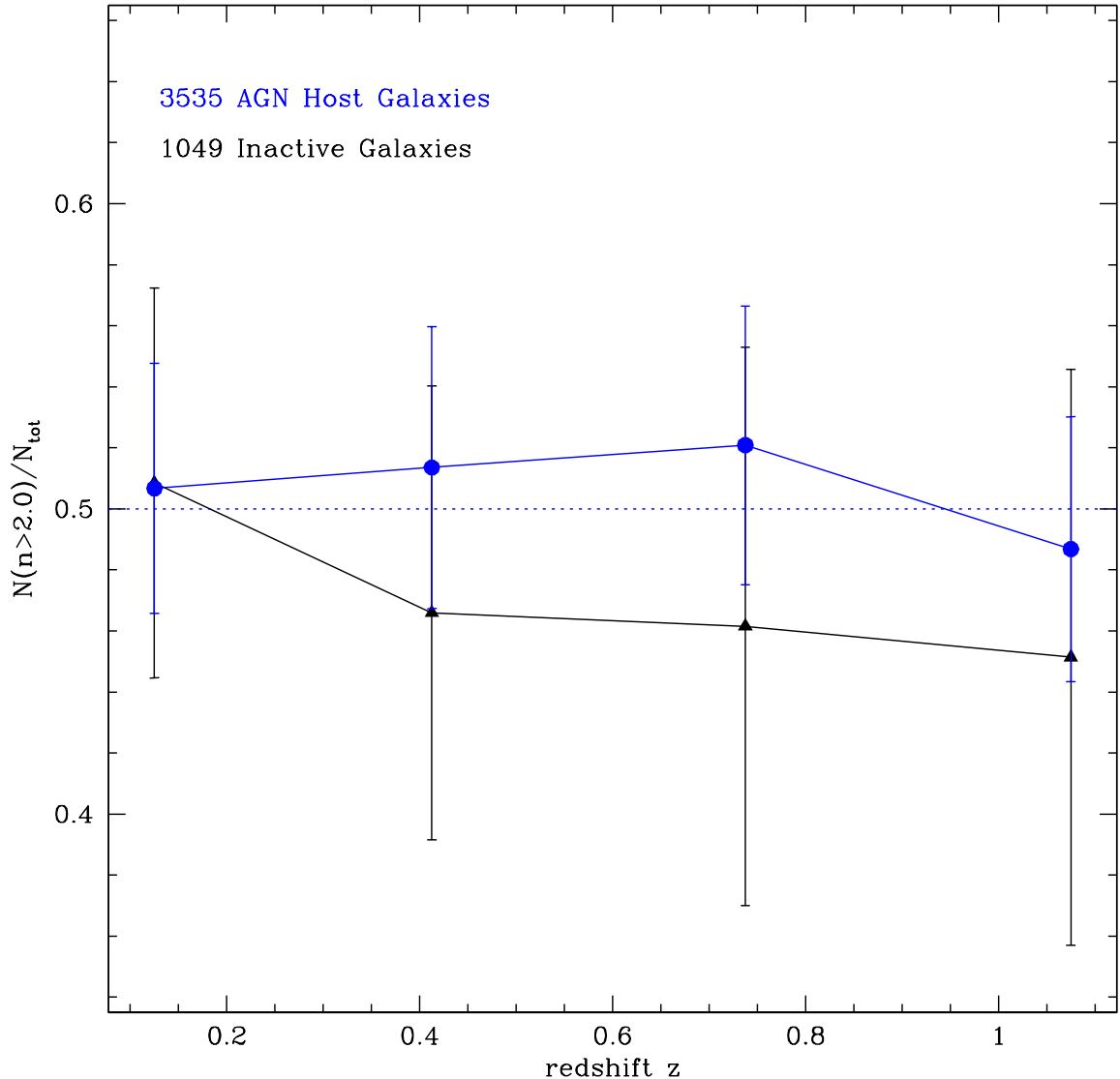}
\caption{ Fraction of galaxies classified as bulges with
redshift. Blue circles represent simulated AGN hosts, and black
triangles represent simulated inactive galaxies (error bars determined
using Poisson statistics). Because the initial simulation contains
equal numbers of pure bulges and disks, the actual bulge fraction is
0.5 for both samples. At the lowest redshift bin, these fractions are
recovered to within 1\% for both samples. The sample of simulated AGN
host galaxies is mis-classified by up to 2\% (0.02), and the inactive
sample is mis-classified by up to 5\%. Uncertainties are higher for
inactive galaxies due to lower numbers of galaxies in the inactive
sample, but in both cases the recovered fractions are consistent with
the correct fraction within the uncertainties.}
\label{nclass_zavg}
\end{figure}

Figure \ref{nclass_zavg} thus shows the fitted morphological fraction
of bulges with redshift of our simulated sample of pure bulges and
disks for both AGN hosts and inactive galaxies. After applying the
radius and magnitude cuts above (i.e., removing only unphysical
fit results from the sample), we calculate the fraction of galaxies in
each redshift bin that would be classified as bulge-dominated
according to an $n > 2$ cut. For a perfectly-recovered sample, this
fraction would be $N(n>2.0)/N_{tot} = 0.5$ at all redshifts because
we simulated equal numbers of bulges and disks. However, the
convergence rate for pure bulge and disk fits is slightly different at
each redshift, and even our set of reasonable and conservative error
cuts removes different fractions of bulge and disk galaxies at each
redshift. While this does affect the morphological composition of the
sample at each redshift, its effect is small: the morphological
fraction of $n>2$ galaxies for AGN hosts is typically within 2\% of
the intrinsic fraction at all redshifts. The inactive galaxy
morphological fraction is within 5\% of the intrinsic fraction at all
redshifts; both samples are consistent within their uncertainties with
the intrinsic fraction.

\begin{figure}
\plotone{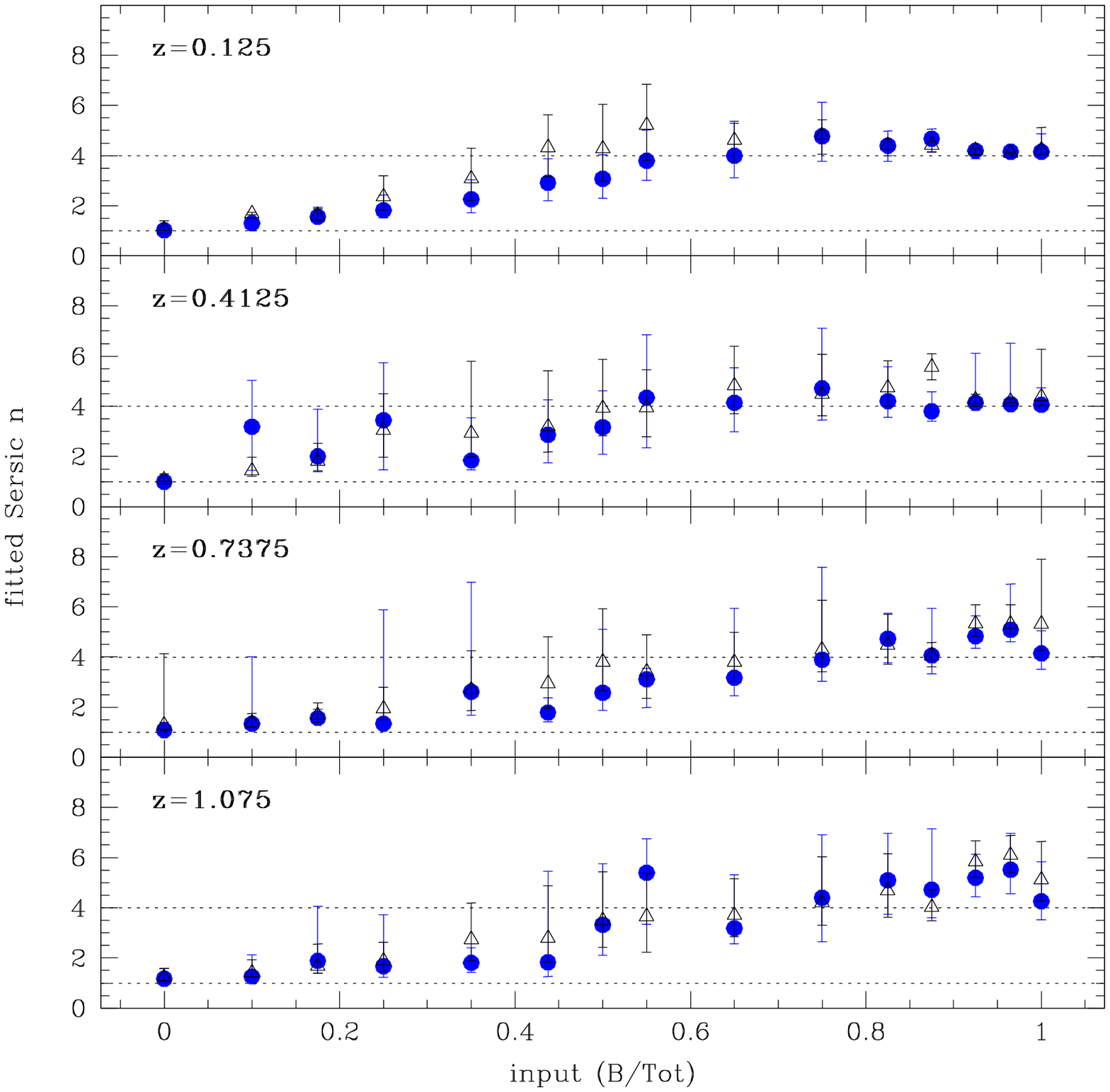}
\caption{Fitted S\'ersic index vs. input \BT\ ratio, for simulated
normal galaxies (black triangles) and galaxies with added point
sources (blue circles). At all redshifts (low to high, top to bottom)
the fit correctly finds disks ($n=1$) at $\mBT = 0$ and bulges ($n=4$)
at $\mBT = 1$. However, some galaxies fit to S\'ersic values of $n
\geq 4$ despite having up to 30\% of their intrinsic light from a
disk. Galaxies with intermediate values of $n$ ($2 < n < 4$) likely
have bulges that contribute between 20\% and 70\% of the total galaxy
light. Points plotted are the median of each bin's distribution; error
bars mark the widths of each distribution enclosing the central 68\%
of sources in the bin.}
\label{n_BDavg_z}
\end{figure}

For the two-component B+D galaxies, the dependence of fitted S\'ersic
index on input \BT\ ratio is shown in Figure \ref{n_BDavg_z}. We see a
general dependence of fitted $n$ on the intrinsic \BT\ ratio. The
relation changes somewhat as the sample is redshifted, but at all
redshifts the S\'ersic index indicates that galaxies may be fitted
with a S\'ersic index indicative of a pure bulge ($n=4$) even in cases
where a disk is present and contributing up to 45\% of the total
galaxy light. However, a fitted S\'ersic index consistent with an
exponential disk ($n=1$) typically indicates very little bulge
contribution ($< 10$\%); galaxies fitted with $n<1.5$ have $\mBT <
0.2$. Intermediate values of the S\'ersic parameter, $1.5<n<3$,
generally indicate a galaxy with both bulge and disk, but the \BT\
ratio may vary between $0.2 \leq \mBT \leq 0.65$, indicating that it
is impossible to determine with a single-component fit whether a host
galaxy with an intermediate S\'ersic index is intrinsically bulge- or
disk-dominated. Given that two-component B+D fits are also uncertain
in the presence of a central point source, we conclude that
determination of the \BT\ ratio is uncertain by at least 20\% in AGN
host galaxies, with a higher uncertainty for hosts with intermediate
S\'ersic indices and/or fitted $\mBT$.

\begin{figure}
\plotone{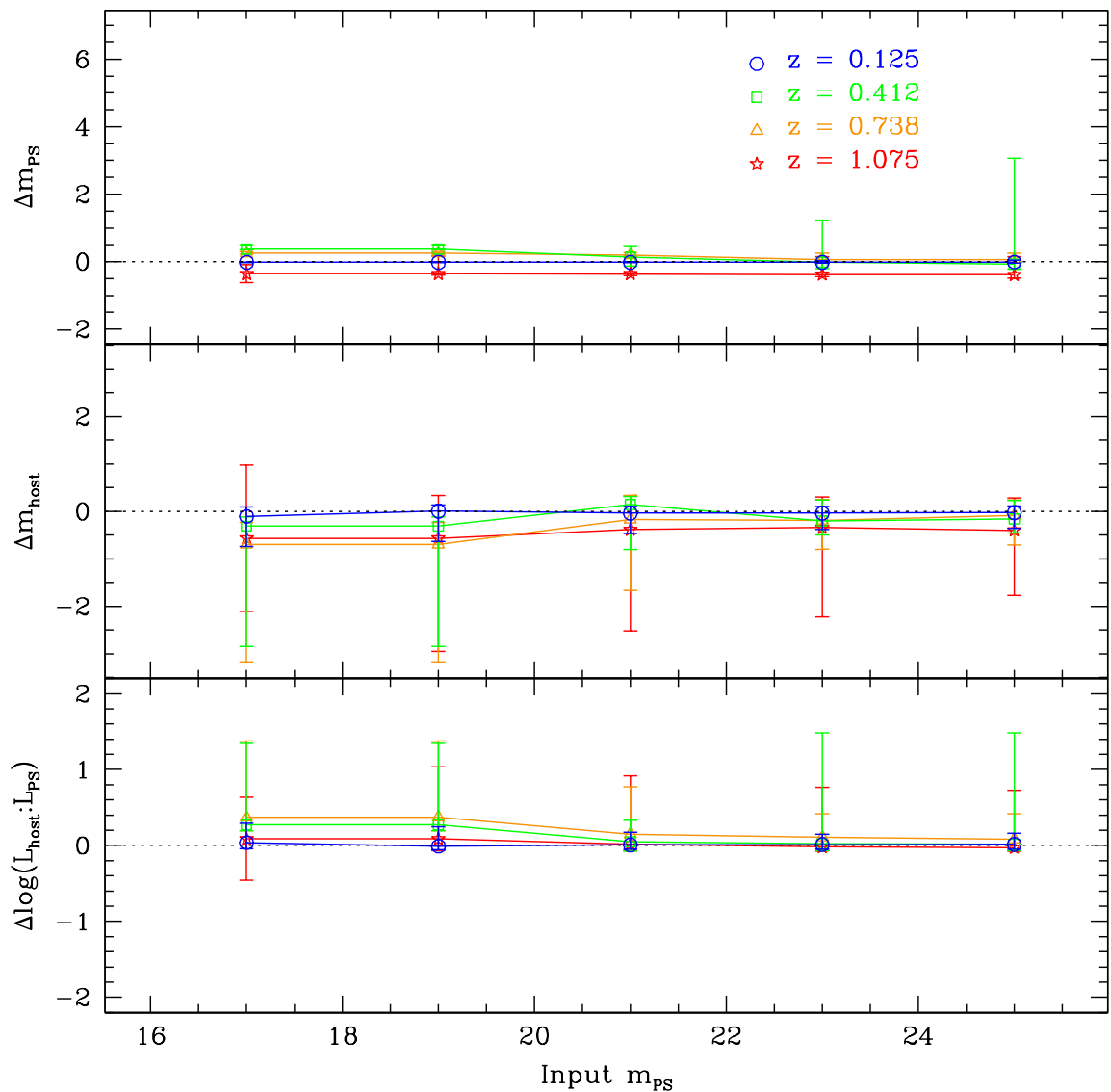}
\caption{ Effects of redshift on recovery of input point sources
(top), input host magnitudes (middle), and \CR\ contrast ratio, shown
for all our simulated host galaxies. Each panel shows the fitted
quantity versus the input point-source magnitude. Each line in
a panel represents a different redshift bin: $z = 0.125$ (blue,
circles), $z = 0.4125$ (green, squares), $z=0.7375$ (orange,
triangles), and $z=1.075$ (red, stars). }
\label{mp_mhost_CR}
\end{figure}

Figure \ref{mp_mhost_CR} shows the recovery of detected input
point sources, hosts, and \CR\ contrast ratios for our simulated
galaxies (B, D, and B+D) at all redshifts. Again we see that for the
local sample, the recovery is excellent: not only are point sources
detected at least 95\% of the time, but the recovered magnitude is
typically within at least $0.1\pm 0.12$ dex of the intrinsic
magnitude. In fact, point-source recovery is very good at all
redshifts: the recovered magnitude is within 0.5 dex of the input
magnitude for all redshifts and input point-source magnitudes.

The dispersion in recovered values for the host galaxy magnitude is
higher than that of point-source magnitude values for all
redshifts. While in the local redshift bin the median recovered host
magnitude is within 0.1 dex of the input magnitude, the central 68\%
of values has a larger spread. At higher redshifts, the host galaxy is
more likely to converge to magnitudes that are too bright compared to
the input magnitude. This effect is more pronounced for sources with a
brighter input point source, as expected. For sources with a bright
point source compared to the host, the difference is $\Delta m_{host}
= 0.6$ for $z > 0.7$. For $z \approx 1$, the distribution of $\Delta
m_{host}$ has an extended tail to negative (too-bright) values: a
significant fraction of high-redshift sources overestimates the host
brightness by up to 1.8 dex regardless of the \CR\ contrast ratio. For
$z < 1$, this tail also exists for all \CR\ values. However, when the
input point source is faint compared to the host, the host magnitude
is fitted to within $-0.8 < \Delta m_{host} < 0.3$ ($1\sigma$) for
these redshifts.

\section{Summary and Conclusions}

We simulated 54,418 GOODS ACS galaxy images of inactive and AGN host
galaxies at redshifts $0.125 < z < 1.25$, the largest sample of simulated
two-dimensional galaxy morphologies to date.

Using a robust initial guess routine followed by batch-fitting, we
performed single-component and bulge-disk host galaxy fits on these
simulated galaxies, while also extracting the central point-source
components of the AGN.

For batch-fit galaxies with central point-source components, typically
60\%-70\% of the fits converge successfully and pass our error cuts. This
percentage can be significantly increased by follow-up case-by-case
fitting, starting from the results of the batch fitting routines. 

We reliably extract central point-source magnitudes for the simulated
AGNs using a simple \csqnu\ cut on the recovered fits. Using a
combination $\mcsqnu < 2$ and relative effective radius error cut of
$\left( \sigma_{r_e} / r_e \right) = 0.8 $ removes 88\% of the poor
host galaxy fits while excluding 12\% of good host galaxy fits. The
accurate extraction of AGN and host galaxy parameters does not depend
on the host:AGN contrast ratio $\contr$, as long as it is greater
than 1:4.

Galaxy size is recovered to within $\Delta r_e < 1$ kpc for
single-component galaxies. For individual bulges and disks within
two-component bulge + disk host galaxy fits, the distribution of
recovered radii peaks within $\Delta r_e < 1$ kpc at all redshifts,
but the large width of the distributions indicates that individual
fits to bulges and disks within two-component galaxies are highly
uncertain.

The average recovered morphological fraction of our sample is within
5\% of the true fraction, even at $z \approx 1$. Of course, this
number should be regarded as a lower limit since actual data samples
of two-dimensional AGN host galaxy morphologies are likely to be
significantly smaller than our simulations, and thus suffer from more
uncertain statistics.  We recover the correct morphology 99\% of the
time for inactive galaxies out to $z = 1.25$.

The average \BT\ ratios are within 10\% of the true ratio at all
redshifts, but with large scatter ($\sim 50$\%). This means that
values for individual hosts are unreliable at the 50\% level, but
statistical population averages are more accurate. Thus, uncertainties
in individual black hole masses determined using $L_{bulge}-M_{BH}$
relations \citep{mclure02, marconi03, ferrarese05} are higher when
using fitted bulge luminosities of AGN host galaxies than those of
inactive galaxies, and may be systematically overestimated due to the
presence of central point sources.

Galaxies with S\'ersic $n < 1.5$ are generally disk-dominated, with at
least 80\% of their total light coming from a disky
component. Galaxies with intermediate $1.5 < n < 3$ have larger bulge
components (occupying from 20\% to 65\% of the total galaxy
light). Galaxies that appear to be bulge-dominated, i.e., with
$n \geq 3$, may derive as little as 45\% (depending on redshift) of
their total light from a bulge component; for $n=4$, typically $\sim
70$\% of the galaxy light comes from the bulge.

In fully simulated AGN host galaxies, the central point source is
correctly detected over 90\% of the time, with only weak dependence on
the intrinsic contrast ratio between the host galaxy and point source
or the host morphology. This indicates that a sample of AGN host
galaxy fits to ACS data with $z < 1.25$ is at least 90\%
complete in its detection of central point sources.

However, we also detect spurious point sources in simulated galaxies
where no central point source is present, for as little as 1\% or as
high as 25\% of the galaxies, depending on the morphological
composition of the sample. Fits to bulge-dominated galaxies are far
more likely to detect a point source when none are present than fits to
disk-dominated galaxies.

\begin{table}
\begin{center}
\begin{tabular}{|c|c|}\hline
Fitted S\'ersic index & Intrinsic bulge-to-total ratio \\ \hline
$n \leq 1.5$ & $\mBTlet < 0.2$ \\
$1.5 < n < 3$ & $0.2 < \mBTlet < 0.65$ \\
$n \geq 3$ & $\mBTlet > 0.45$ \\
$n \geq 4$ & $\mBTlet > 0.55$ \\ \hline

\end{tabular}
\caption{ Relationship of a host galaxy's fitted S\'ersic index to its
intrinsic \BT\ light ratio. Hosts with intermediate S\'ersic index may
have a wide range of \BT\ ratios; hosts with fitted $n=4$, typically
classified as pure deVaucouleur bulges, may in fact have \BT\ ratios
as low as 55\%.}
\label{res_summ}
\end{center}
\end{table}

We have used these simulations to evaluate the robustness of our
results for AGN host galaxies observed with ACS. This is important for
assessing the accuracy of recovered host morphologies, as well as the
evolution of AGN host morphology with redshift. All of our results
that can be quantitatively applied to a sample of real AGN host
galaxies are also presented in tabular format (Tables
\ref{rms_params}, \ref{rms_params_CR2}, and \ref{res_summ}).

We have performed these simulations using the image depths and noise
properties of GOODS, but they may also be used for other ACS surveys
such as GEMS \citep{rix04}, COSMOS \citep{scoville07}, and AEGIS
\citep{davis07}. The extent to which these simulations are directly
applicable or provide limits on the accuracy of AGN host galaxy
morphological fits for other ACS data depends on the relative depth of
observations in each survey.

It is important to note that these results apply to redshifts where
the ACS bands are observing rest-frame optical data, e.g. $z <
1.25$. As we move into the era of newer telescopes such as the
\emph{James Webb Space Telescope} and observing rest-frame optical
wavelengths at very high resolution becomes possible for higher
redshifts, an understanding of the effects of a central point source
on high-redshift morphologies will become crucial. Simulations to
probe these effects have begun on small scales \citep{dasyra08}, but
more such work is needed.


\acknowledgments

The authors wish to thank C. Conselice, L. Moustakas, P. Natarajan,
S. Ravindranath, and the entire GOODS team for helpful discussions
that improved this work. Thanks especially to C. Peng for making
GALFIT publicly available and for many enlightening discussions. We
acknowledge support from NASA through grants HST-AR-10689.01-A,
HST-GO-09425.13-A and HST-GO-09822.09-A from the Space Telescope
Science Institute, which is operated by the Association of
Universities for Research in Astronomy under NASA contract NAS
5-26555.

\bibliographystyle{apj}
\bibliography{refs}
\end{document}